\documentclass[aps,prd,twocolumn,superscriptaddress,nofootinbib]{revtex4}
\usepackage{graphicx}
\usepackage{amsmath,amssymb,latexsym}
\usepackage{bm}
\usepackage{epsfig}
\newcommand{\postscript}[2]{\setlength{\epsfxsize}{#2\hsize}
   \centerline{\epsfbox{#1}}}

\newcommand{\nep}{n(\epsilon)}
\newcommand{\En}{E_n}
\newcommand{\Fn}{F_n}
\newcommand{\Enu}{E_{\overline{\nu}}}

\newcommand{\Enubar}{\epsilon_{\overline{\nu}}}
\newcommand{\thnu}{\overline{\theta}_{\overline{\nu}}}
\newcommand{\cth}{\cos\overline{\theta}_{\overline{\nu}}}
\newcommand{\Fnu}{F_{\overline{\nu}}}
\newcommand{\tbar}{\overline{\tau}_n}

\newcommand{\Enmax}{E_n^{\rm max}}

\newcommand{\eps}{\epsilon}

\newcommand{\nutau}{\nu_\tau}
\newcommand{\nue}{\nu_e}
\newcommand{\numu}{\nu_\mu}

\begin{document}

\title{TeV $\bm{\gamma}$-rays and neutrinos from photo-disintegration
  of nuclei in Cygnus OB2}
\author{Luis A.~Anchordoqui}
\affiliation{Department of Physics, University of Wisconsin-Milwaukee,
P.O. Box 413, Milwaukee, WI 53201
}

\author{John F. Beacom}
\affiliation{CCAPP, Departments of Physics and Astronomy,
Ohio State University,
Columbus, OH 43210
}

\author{Haim Goldberg}
\affiliation{Department of Physics,
Northeastern University, Boston, MA 02115
}

\author{Sergio Palomares-Ruiz}
\affiliation{Department of Physics and Astronomy,
Vanderbilt University, Nashville, TN 37235
}
\affiliation{Institute for Particle Physics Phenomenology,
University of Durham, Durham DH1 3LE, UK}

\author{Thomas J. Weiler}
\affiliation{Department of Physics and Astronomy,
Vanderbilt University, Nashville, TN 37235
}

\begin{abstract}
  \noindent TeV $\gamma$-rays may provide significant information
  about high energy astrophysical accelerators. Such $\gamma$-rays can
  result from the photo-de-excitation of PeV nuclei after their parents
  have undergone photo-disintegration in an environment of ultraviolet
  photons. This process is proposed as a candidate explanation of the
  recently discovered HEGRA source at the edge of the Cygnus OB2
  association.  The Lyman-$\alpha$ background is provided by the rich
  O and B stellar environment.  It is found that (1) the HEGRA flux
  can be obtained if there is efficient acceleration at the source of
  lower energy nuclei; (2) the requirement that the Lorentz-boosted
  ultraviolet photons can excite the Giant Dipole Resonance implies a
  strong suppression of the $\gamma$-ray spectrum compared to an
  $E_\gamma^{-2}$ behavior at energies $\alt 1~{\rm TeV}$
  (some of these energies will be probed by the upcoming GLAST mission);
  (3) a TeV neutrino counterpart from neutron decay following
  helium photo-disintegration will be observed at IceCube only if a
  major proportion of the kinetic energy budget of the Cygnus OB2
  association is expended in accelerating nuclei.
\end{abstract}


\maketitle

\section{Introduction}

There are two well-known mechanisms for generating TeV $\gamma$-rays in
astrophysical sources~\cite{Aharonian:2004yt}. The first involves
purely electromagnetic (EM) processes, including synchrotron emission
and inverse Compton scattering.  The second may be termed hadronic, in
which the $\gamma$-rays originate in $\pi^0$ decays. The latter can in
turn be traced to either $pp$ or $p\gamma$ collisions.

The EM processes originate in the acceleration of electrons; as a
result, the TeV $\gamma$-rays must be accompanied by an X-ray
counterpart. In addition, the conflicting requirements on the magnetic
field (large field for acceleration, small field for limiting
synchrotron cooling) limits the energy of the emitted photons
$E_\gamma\alt 10$ TeV.

The hadronic $p\gamma$ mode is characterized by a large threshold for
$\pi^0$ production, and thus it is favored only in very hot photon
environments, or in the presence of very energetic proton beams. In
addition, the presence of the high threshold allows acceleration free
of scattering losses as long as the Hillas
criterion~\cite{Hillas:1985is} holds. In the $pp$ hadronic mode,
threshold effects are insignificant. Because of this, it is generally
assumed that $\pi^0$ production occurs in a region distinct from the
site of the primary acceleration. However, one can set conditions
relating the interaction length of the nucleons, the neutron decay
lifetime and the source confinement radius to permit simultaneous
emission of neutrons, $\gamma$-rays and
neutrinos~\cite{Ahlers:2005sn}.  In the $p\gamma$ mode the spectrum of
$\gamma$-rays follows that of the parent proton population in the
energy region above threshold. In the $pp$ mode, the resulting
$\gamma$-ray spectrum is broad, reflecting the presence of a
quasi-Feynman plateau which spans the entire rapidity space. The
photon spectrum will deviate from the proton spectrum if the Feynman
plateau is not flat in rapidity space~\cite{Anchordoqui:2004bd}, as
suggested by Tevatron data~\cite{Abe:1989td}. Finally, for both the
$p\gamma$ and $pp$ modes, the decay of the charged pions yield a
neutrino counterpart with energy and intensity similar to the photons.

In this paper, we discus in detail a third dynamic which can lead to
TeV $\gamma$-rays: the photo-disintegration of nuclei at the source,
followed by the photo-de-excitation of the daughter
nuclei~\cite{Moskalenko:phd}. In order to generate TeV $\gamma$-rays
as a result of emission of MeV $\gamma$-rays in the rest frame of the
de-exciting nucleus, the Lorentz factor of the boosted nucleus must be
$\sim 10^6.$ For this boost factor, excitation via the Giant Dipole
Resonance ($\sim 10~{\rm MeV} - 30~{\rm MeV}$ in the nucleus rest
frame) is obtained with ambient photons with energies in the far
ultraviolet (usually defined as 1-20~eV).
Photons of these energies are expected from the Lyman
$\alpha$ emissions from hot stars.  This process clearly reduces the
threshold energy requirement relative to $p\gamma$.  The important
role played by the Giant Dipole Resonance (GDR) in the
photo-disintegration effectively suppresses the $\gamma$-ray spectrum
below 1~TeV.

\begin{figure}
\begin{center}
\postscript{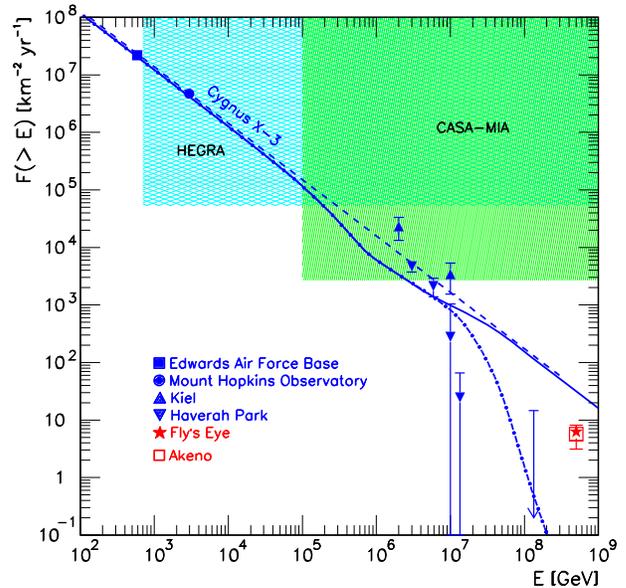}{0.98}
\caption{Integrated flux from the binary system Cygnus X-3 as reported
by the different collaborations.  The dotted line is a best single
power law fit, $F_\gamma (> E_\gamma) = 3.3 \times 10^{-8}
\,\,(E_\gamma/{\rm GeV})^{-0.98} \, \,\, {\rm cm}^{-2}\,\, {\rm
s}^{-1},$ to the Cygnus X-3 data: $\blacktriangle$~\cite{Samorski},
$\blacksquare$~\cite{Danaher}, $\bullet$~\cite{Danaher}, and
$\blacktriangledown$~\cite{Lloyd-Evans:nx}.  The accompaning solid and
dash-dotted lines take into account attenuation on the CMB, with the
latter also including a cutoff at the source. Also shown are the
integrated fluxes of neutral particles from the direction of Cygnus
X-3 reported by the Akeno ($\Box$~\cite{Teshima:1989fc}) and Fly's Eye
($\star$~\cite{Cassiday:kw}) collaborations. The cross-hatched bands
indicate the 90\%~CL upper limit on steady state flux from the
direction of Cygnus X-3 as observed by the
CASA-MIA~\cite{Borione:1996jw} and HEGRA~\cite{HEGRA05} experiments.}
\label{cygnusx3}
\end{center}
\end{figure}

In recent decades, the Cygnus Spiral Arm has been a site of
$\gamma$-ray signal candidates which tend to come and go. In the
energy band $2 \times 10^6 < E_\gamma/{\rm GeV} < 2 \times 10^7,$ data
collected by the Kiel air shower experiment~\cite{Samorski} show a
$4.4\sigma$ excess of events in the direction of the binary system
Cygnus X-3, with the typical 4.8 hr modulation previously observed in
the MeV~\cite{Lamb} and TeV~\cite{Danaher} regions.  This result,
including the X--ray binary period, was later confirmed by the Haverah
Park data~\cite{Lloyd-Evans:nx}, with the additional observation of an
abrupt steepening of the spectrum at $E_\gamma > 2 \times 10^7~{\rm
GeV}$. The distance ($\approx 10$~kpc) to Cygnus X-3 slightly exceeds
the minimum path length for cosmic microwave background (CMB)
absorption in the energy band of the Kiel
experiment~\cite{Anchordoqui:2002hs}. However, since absorptive effects
decrease for energies beyond $10^7$~GeV (see Fig.~\ref{cygnusx3}), the
observed steepening must be traced to a softening of the injection
spectrum.

At a much higher energy $\agt 5 \times 10^8~{\rm GeV},$ the analysis
of the cumulative Fly's Eye data also revealed an excess of events
from the direction of Cygnus X-3, with chance probability of $6.5
\times 10^{-4}$~\cite{Cassiday:kw}. A $3.5\sigma$ excess from this
direction and in the same energy region has also been observed at the
Akeno air shower array~\cite{Teshima:1989fc}.  The inferred signal
fluxes are consistent at the 1$\sigma$ level.  However, the evidence
for the 4.8 hr modulation is termed as ``weak'' by the Fly's Eye
Collaboration, and is absent in the Akeno data. Thus, one may infer
that the high energy signals originate in a different source within
the angular field of view of Cygnus X-3.

More recent data from the CASA-MIA~\cite{Borione:1996jw} and the
HEGRA~\cite{HEGRA05} experiments place restrictive bounds on steady
state fluxes from the Cygnus region. These can be seen in
Fig.~\ref{cygnusx3}, and  suggest that the earlier reported
fluxes do not reflect current steady state activity. At this point, it
is critical to note that only the HEGRA experiment has the angular
resolution to place an upper limit on the steady state flux from
Cygnus X-3, and at the same time observe significant activity (at the
7$\sigma$ level) in the TeV region from an unidentified source (with
no optical or X-ray counterparts) which is within $0.5^{\circ}$ of the
X-ray binary~\cite{HEGRA05}. An excess at the $3.3\sigma$ level from
the direction of this unidentified source is also present in the
Whipple data~\cite{Lang:2004bk}. The strength and specificity of the
source, and the distinct absence of an X-ray counterpart, make this
source a good candidate for probing the nucleus
photo-disintegration/de-excitation model for producing TeV $\gamma$-rays.

Especially intriguing is the possible association of the TeV HEGRA
source with Cygnus OB2, a cluster of several thousands of young hot OB
stars. At a relatively small distance ($\approx 5000$~light years) to
Earth, this is the largest massive Galactic stellar association.
Cosmic ray nuclei are expected to be trapped and accelerated through
turbulent motions and collective effects of star winds. In this paper
we present a model for explaining the HEGRA observations, in which the
trapped high energy nuclei undergo stripping on the starlight
background and their surviving fragments emit $\gamma$-rays in
transition to their ground states. A short version highlighting the
salient features of the model has been issued as a companion
paper~\cite{shortP}. The outline here is as follows: in Sec.~\ref{II}
we provide a detailed description of the salient characteristics of
the Cygnus OB2 association, including stellar counts, lifetime, size
and energy considerations. Section~\ref{III} contains a description of
the HEGRA TeV source, and summarize the different proposed
explanations. The model is presented in Sec.~\ref{IV}, including a
calculation of the TeV photon flux, following a discussion of the
accompanying neutrino flux. In Sec.~\ref{V} we compare the TeV
$\gamma$-ray and neutrino yields in the model with those corresponding
to the hadronic modes. In Sec.~\ref{VI} we examine the possibility
whether our model can accommodate the emission of ultra-high energy
cosmic rays. Conclusions are collected in Sec.~\ref{VII}.

\section{Cygnus OB2 in a Nutshell}
\label{II}

The Cygnus OB2 (VI Cygni) association is one of the most massive
associations, with some of the most luminous stars, in our Galaxy. It
was first noticed by M\"unch and Morgan~\cite{MM53} who, during the
course of a classification of blue giants (OB stars), found eleven of
these objects in that region. Since the pioneering spectroscopy and
photometry of Johnson and Morgan~\cite{JM54} and subsequent
works~\cite{early}, this region has been known to harbor a large
population of massive and early type stars, which have been found to
be highly reddened. The first comprehensive study of this stellar
association~\cite{LR65,RLP66,RLP67} identified a few hundred OB stars
as possible members based on photographic photometry. Other photometric
and spectroscopic studies~\cite{other} were carried out during
following years confirming and extending the first results.

Cygnus OB2 is located at galactic coordinates $(l,b) \sim (80^o,1^o)$,
behind the Great Cygnus Rift. The study performed in
Ref.~\cite{RLP66}, inferred an elliptical shape with major and minor
axes of 48' and 28', respectively, and estimated more than 3000 stars
with at least 300 of OB spectral type, which resulted in a total mass
for the association of $(0.6-2.7) \times 10^4 M_{\odot}$. However, due
to the extreme redenning in and around this region, which hampers the
detection of even OB stars, the observed morphology was rather an
artifact created by the particular extinction pattern of the field. A
more recent work~\cite{K00} based on a statistical study of point
sources revealed by the {\sl Two Micron All Sky Survey} (2MASS) in the
near-infrared, shows that this association is much larger and richer
than previously thought. The resulting stellar distribution reveals a
rather regular and almost circular density profile with the center
located at $(\alpha, \delta) = (20^h 33^m 10^s, +41^o 12')$ and with a
pronounced maximum slightly offset at $(\alpha, \delta) = (20^h 33^m
10^s, +41^o 15.7')$. Stars counts show that 50\% of the members are
located within a radius of $21'$, and 90\% within a radius of $45'$
around the center, merging with the field stars at a radius of $\sim
1^o$. The central stellar density reaches 4.5 stars/arcmin$^2$ above
the field star density, and drops to 50\% at a radius of $13'$. By
integrating the radial density profile, after substraction of the
field star density, the total number of OB stars was found to be $2600
\pm 400$~\cite{K00}, much larger than the earlier estimate of
$\sim$300~\cite{RLP66}. Furthermore, the total number of O-type stars
was inferred to be $120 \pm 20$~\cite{K00}. Nevertheless, this number
must be cast as an upper limit, as the 2MASS data do not allow a
precise spectral determination and hence do not account for possible
evolutionary effects, i.e., the quoted number reflects the
initial number of very massive ($\agt 20 M_\odot$) stars, although
some may have already evolved. This also agrees with more recent
estimates~\cite{Comeron02,H03} and suggests a total mass for the
association of $(4-10) \times 10^4 M_\odot$, where the primary
uncertainty comes from the unknown lower mass cut-off. Using the
radial density profile, a central mass density of $(40-150) M_\odot /
{\rm pc}^3$ is determined~\cite{K00}.

Early distance determinations revealed the proximity of this region,
varying from 1.5 kpc~\cite{JM54} to 2.1~kpc~\cite{RLP66}. More recent
estimates have set this distance to $d \sim 1.7~{\rm
kpc}$~\cite{TTC91,MT91}.  At such distance, the inner $21'$, with half
the total number of objects, results in a physical radius of $R_{\rm in}
\sim 10~{\rm pc}$, with $R_{\rm out} \sim $~30 pc being the radius of the
association. Yet, a revision of the effective temperature scale of
O-type stars yields a closer distance of $\sim$1.5 kpc~\cite{H03},
when the age of the association is taken into account in fitting the
late-O and early-B dwarfs to model isochrones. It is from the
superposition of the isochrones (calculated using the theoretical
evolutionary tracks of the model of Ref.~\cite{isochrone}) on the
Hertzsprung-Russell diagrams, that the age of the association has been
estimated to lie between 1--4 Myr~\cite{K02}, compatible with other
estimates~\cite{ABC81,TTC91,HCVM99,H03}. This range reflects the
dispersion of the upper main sequence and agrees with the fact that
the observed large number of O-type stars implies that the association
should be younger than $\sim$5~Myr, because in the case of coeval star
formation, the number of this type of stars decreases
rapidly~\cite{K02}. In addition, the fact that there are some
Wolf-Rayet stars within Cygnus OB2 suggests an age larger than $\sim$2
Myr. On the other hand, the non-detection of any supernova remnant in
Cygnus OB2~\cite{WHL91} points to an association younger than $\sim$4
Myr. Nevertheless, the presence of a few slightly evolved supergiants
(an O3 If$^*$ star, three Wolf-Rayet stars and two LBV candidates)
suggests that the star formation in the asociation was not strictly
coeval and there are even indications of ongoing star
formation~\cite{TTC91,MT91,PJB92,PK98,HCVM99}. This fact would narrow
down the quoted age boundaries.

Early estimates of the wind mechanical luminosity for the Cygnus OB2
association gave $L_w \simeq 10^{38} \, {\rm erg \; s}^{-1}$
\cite{ABC81}. However, the improved data on the population of the
region required a revision of this estimate. By using a detailed
analysis of radiatively wind models for hot stars, it has been shown
than at an early stage ($\alt 2$ Myr), the wind mechanical luminosity
is maintained approximately constant~\cite{LRD92}, being $\sim 2
\times 10^{34} \, {\rm erg \; s}^{-1}$ per solar mass, which, for the
association, yields $L_w \simeq (1-2) \times 10^{39} \; {\rm erg \,
  s}^{-1}$ \cite{LPF02}. After the first $\sim 2$ Myr, the luminosity
increases further as Wolf-Rayet stars appear. This estimates assumes
stars of solar metallicity and a Salpeter initial mass function (IMF)
slope. There have been different calculations for the IMF
slopes~\cite{MT91,PJB92,K00,K02} which however do not agree with each
other, although lie in the interval $\Gamma \simeq -(1.0 - 1.6)$
($\Gamma_S = -1.35$ being the Salpeter slope). Nevertheless, these
discrepancies are due to systematic uncertainties and hence, a
reasonable approximation is to assume a uniform canonical Salpeter
slope. Should the metallicity be close to the solar value, this
implies~\cite{LRD92} that the estimate of the wind mechanical
luminosity is probably correct within a factor of $\sim 2$. Other
values for the metallicity would modify this prediction by at most a
factor $\alt 10$. On the other hand, the total Lyman continuum
luminosity of the stars has been estimated to be $\sim 10^{51} {\rm
photons \  s}^{-1}$~\cite{K02}.

Recently, there have been observations of TeV $\gamma$-rays from the
northeast boundary of this association (see
Fig.~\ref{cygOB2})~\cite{HEGRA05}. These observations will be
discussed in the following section, and will constitute the focus of
this paper.

\section{T\lowercase{e}V J2032+4130}
\label{III}

The HEGRA system of Imaging Atmospheric \v{C}erenkov Telescopes (HEGRA
IACT-system) consisted of five identical \v{C}erenkov telescopes (each
with 8.5 m$^2$ mirror area) and employed a stereoscopic technique
achieving an angular and energy resolution better than $0.1^o$ and
15\%, respectively, for $\gamma$-rays on an event-by-event basis with
energies from 0.5~TeV to $\sim$ 50~TeV~\cite{HEGRA}.

The observation of the Cygnus region by the HEGRA IACT-system has
allowed the serendipitous discovery of a TeV source in the outskirts
of the core of Cygnus OB2~\cite{HEGRA02}. The analysis of the total
278.3 hours of observations performed in two periods from 1999 to 2002
(120.5 hours from 1999 to 2001~\cite{HEGRA02} and 157.8 hours during
2002~\cite{HEGRA05}) has revealed the presence of a steady
(and possibly extended) TeV source, with
hard injection spectrum. Interestingly, there
have been earlier claims of a multi-TeV excess in this
region~\cite{TeV}.

The excess significance of the TeV source is $7.1\sigma$ and it
appears extended at more than $4\sigma$ level with a morphology which
is suitably described by a Gaussian profile. The source is termed TeV
J2032+4130 after the position of the center of gravity.
Its extension (Gaussian $1\sigma$ radius) is $6.2'(\pm1.2'_{\rm stat} \pm
0.9'_{\rm sys})$, which at a distance of 1.7 kpc results in $r \sim
3.07 (\pm 0.59_{\rm stat} \pm 0.45_{\rm sys})$ pc.

Three different types of intrinsic morphology were
tested~\cite{HEGRA05} (disc, volume and surface), but the data cannot
discriminate between them. The energy spectrum determination yielded a
pure power-law fit with a hard photon index, showing no indication for
an exponential cut-off, given by~\cite{HEGRA05}
\begin{widetext}
\begin{equation}
\label{Hflux}
\frac{dF_\gamma}{dE_\gamma}  =  6.2 \, (\pm1.5_{\rm stat} \, \pm \,
1.3_{\rm sys}) \times 10^{-13} \left(\frac{E_\gamma}{{\rm
    TeV}}\right)^{-1.9(\pm0.1_{\rm stat} \pm 0.3_{\rm sys})} {\rm
  cm}^{-2} {\rm s}^{-1} {\rm TeV}^{-1}  \equiv
     N_{\rm HEGRA} \, \left(\frac{E_\gamma}{{\rm
      TeV}}\right)^{-\alpha_{_{\rm HEGRA}}} \,,
\end{equation}
\end{widetext}
which implies a flux above 1 TeV given by
\begin{equation}
\label{Hfluxint}
F_\gamma(E_\gamma>1 {\rm TeV}) = 6.9 \, (\pm 1.8_{\rm stat}) \times
  10^{-13} {\rm cm}^{-2} {\rm s}^{-1}.
\end{equation}
These results imply a luminosity of $\sim 10^{32}$ erg/s above 1~TeV,
which is well within the kinetic energy budget of Cygnus OB2, and
indeed, also within that of a number of notable member
stars~\cite{MT91,stars}.

\begin{figure}
\begin{center}
\postscript{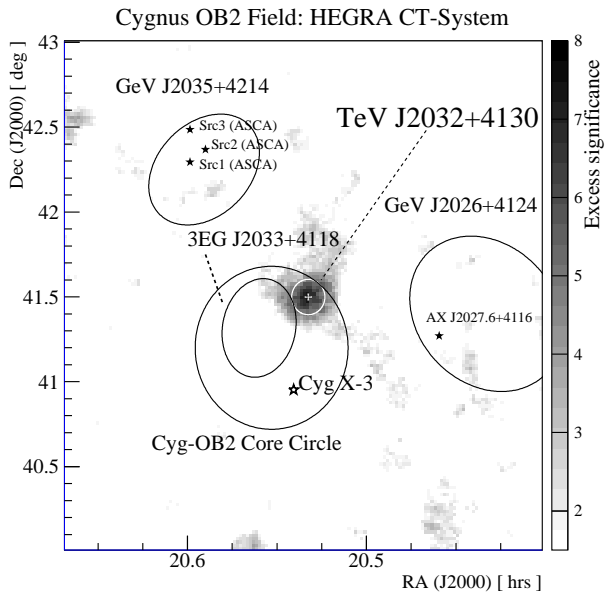}{0.98}
\caption{Skymap of correlated event excess significance ($\sigma$)
from all HEGRA IACT-System data ($3^\circ \times 3^\circ$ FoV)
centered on the TeV source J2032+4130. Nearby objects are also shown:
95\% contours for 3 EGRET sources (indicated by the ovals), their
possible X-ray associated counterparts (as given in
Ref.~\cite{Roberts:2000zr}), and Cygnus X-3. The center of gravity
with statistical errors and intrinsic size (standard deviation of a
2-dim Gaussian, $\sigma_{src}$) are indicated by the white cross and
white circle, respectively. The TeV source, J2032+4130, is positioned
at the edge of the error circle of the EGRET source 3EG J2033+4118,
and within the core circle of the extremely dense OB stellar
association Cygnus OB2~\cite{HEGRA05}.}
\label{cygOB2}
\end{center}
\end{figure}

As can be seen in Fig.~\ref{cygOB2}, this signal is located at the
edge of the error circle of the EGRET source 3EG J2033+4118 and within
the $\sim$10 pc-radius core circle of the Cygnus OB2 association. The
EGRET observations~\cite{Hartman:1999fc} lie in the energy range below
10~GeV, and closely display a spectrum $\propto E_\gamma^{-2}.$ It
will be of important significance for our discussion that when
extrapolated to the TeV range according to the $E_\gamma^{-2}$
behavior, the flux exceeds the HEGRA flux by several orders of
magnitude. Hence we surmise that these observations constitute two
different sources.

An additional set of observations performed during 1989-90 by the
Whipple Observatory atmospheric \v{C}erenkov imaging
telescope~\cite{Lang:2004bk} has been recently reanalyzed in the light
of the HEGRA data. These confirm an excess in the same direction as
J2032+4130, although with considerably larger flux, above a peak
energy energy response of 0.6~TeV. The statistical significance of the
signal is only 10\% smaller with selection of events above
1.2~TeV. However, the large differences between the flux levels cannot
be explained as errors in estimation of the sensitivity of the
experiments since they have been calibrated by the simultaneous
observations of other TeV sources.

So far no clear counterparts at other wavelengths have been
identified, and moreover, the observed spectrum is not easily
accommodated with synchrotron radiation by
electrons~\cite{HEGRA02}. The difficulty to accommodate the spectrum
by conventional electromagnetic mechanisms has been exacerbated by the
failure of CHANDRA and VLA to detect X-rays or radiowaves signaling
acceleration of any electrons~\cite{Butt03}.

Nevertheless, a leptonic origin is not yet excluded, especially if the
TeV source is actually located in the vicinity of Cygnus X-3. The
suggested model~\cite{HEGRA02} proposes a jet-driven termination shock
at the boundary where the relativistic jet meets the interstellar
medium. This results in synchroton and TeV inverse-Compton emission
from the accelerated electrons~\cite{AA98}, giving rise to the
observed signal. Such a jet could emanate either from an as-yet
undiscovered microquasar or from Cygnus X-3. Interestingly, the TeV
signal aligns well with the northern error cone of the bi-lobal jet of
Cygnus X-3~\cite{quasar}. For the observed angular separations, this
would place the TeV source at a distance of 10 kpc from Earth, and
around 70 pc from Cygnus X-3.

The fact that there is no catalogued X-ray sources within the
$2\sigma$ error circle of the TeV source would disfavor a leptonic
origin at Cygnus OB2.  We now briefly discuss some models in which
accelerated protons and nuclei interacting with a local dense gas
cloud~\cite{Butt03} or with stellar winds~\cite{TDR04}, produce
photons from $\pi^0$ decay. In the case of gas cloud
interactions~\cite{Butt03}, an interstellar density of $n_{\rm tot}
\sim 30 \, {\rm cm}^{-3}$ and an injection efficiency (the ratio of
cosmic ray energy to the kinetic energy of the accelerating winds) of
0.08\% reproduces well the observed TeV spectrum.  However, such a
large density has been called into question~\cite{TDR04} because a
re-assessment~\cite{Yao:2003rm} of the CO-H$_2$ conversion factor for
this type of environment (used for calibration) would imply a
considerably lower density, $n_{\rm tot} \sim 0.1\, {\rm
  cm}^{-3}$. With such a density, a much larger injection efficiency
$\sim$ 25\% would be required. A different model in terms of cosmic
ray illumination of stellar winds has been suggested to explain the
signal through hadronic interactions in the innermost parts of the
winds of massive OB stars~\cite{TDR04}. However, a more recent
detailed study of the stellar winds~\cite{DT0510}, including pressure
effects and a more accurate estimate of the (lower) mass-loss rate for
OB stars, resulted in a significant reduction in the photon emission
rate. Thus, if only the currently known stars exist, the predicted
flux from this model is too low to comfortably explain the TeV source.

Another suggested possibility has been that of a point-like origin of
the TeV signal related to an unusual transient X-ray source which lies
$7'$ from the center of gravity of TeV
J2032+4130~\cite{MHGEM03}. However, the position and variability of
this X-ray source and mainly, the confirmed extension of the TeV
source makes this a remote alternative, although it remains possible
that several point-like TeV sources could be masquerading as a single
extended source.

The TeV signal could also be interpreted as originated in the wind
nebulae of an as-yet undetected recent pulsar (similar to Vela-type
pulsars), as a result of hadronic or leptonic
interactions~\cite{B03}. This pulsar must have been formed $\sim 10^4$
years ago in a core collapse of one of the massive stars in the Cygnus
OB2 association. However, the non-observation of any supernova remnant
in the region, as well as the greatly lowered estimates of the ambient
gas density,  seem to pose a serious challenge for this model.

In summary, there is no single compelling explanation for the various
characteristics of TeV J2032+4130~\cite{Butt:2006js}. In the next
section, we present a new mechanism for explaining the HEGRA data.

\section{The model}
\label{IV}

There are two processes by which the nucleus may produce
$\gamma$-rays, photo-disintegration of the nucleus by ambient photons
followed by de-excitation of a daughter nucleus, and photo-pion
production followed by decay of the neutral pions. The two processes
have different thresholds and different signatures. In this section, we
are concerned with the former mechanism, emphasized more than a decade
ago by Moskalenko and collaborators~\cite{BKM87}, but largely ignored
by the rest of the $\gamma$-ray community. In the following section
we will compare and contrast the two processes.

\subsection{Nucleus Photo-Disintegration}

The interaction between photons and high energy nuclei results in
the emission of nucleons. The relevant photonuclear interaction
process for the relevant energies have been studied from the point of
view of the collective and the shell models~\cite{photoreview}.
It is shown that collective nuclear states dominate the interaction,
with low angular momentum modes preferred.  In the energy region which
extends from threshold for single-nucleon emission $\sim$~10~MeV up to
$\sim$ 30 MeV the GDR dominates. The GDR typically de-excites by the statistical
emission of a single nucleon. Above the GDR region, and up to the
photo-pion production threshold at $E_{\rm
  th}^\pi=m_\pi\,(1+m_\pi/2m_N)\simeq 145$~MeV, the non-resonant
processes provide a much smaller cross section with a relatively flat
dependence on energy.

The photo-disintegration rate for a highly relativistic nucleus with
energy $E=\gamma A m_N$ (where $\gamma$ is the Lorentz factor)
propagating through an isotropic photon background with energy $\epsilon$ and
spectrum $n(\epsilon)$, normalized so that the total number of photons
in a box is $\int \nep d\epsilon$, is~\cite{S69}
\begin{equation}
R_A  = \frac{c}{\lambda_A} =
\frac{1}{2} \, \int_0^{\infty} \frac{n(\epsilon)}{\gamma^2 \epsilon^2}
\, d\epsilon \, \int_0^{2\gamma \epsilon} \epsilon' \,
\sigma_A(\epsilon') \, d\epsilon' \, ,
\label{rate}
\end{equation}
where $\sigma_A (\epsilon')$ is the cross section for
photo-disintegration of a nucleus of mass $A$ by a photon of energy
$\epsilon'$ in the rest frame of the nucleus.

The cross section for all the different nuclear species has been
obtained through a direct fit to data~\cite{PSB76}. For medium and
heavy nuclei $(A\ge 30)$ the total photon absorption cross section can
be approximated by a dipole form (sometimes called a ``Breit-Wigner'' or ``Lorentzian'')
\begin{equation}
\sigma_A (\epsilon') = \sigma_0 \,\, \frac{\epsilon'^2 \,
\Gamma^2}{(\epsilon'^2_0 - \epsilon'^2)^2 + \epsilon'^2\, \Gamma^2} \,,
\end{equation}
where $\Gamma$ is the width, $\epsilon'_0$ is the central value of the GDR
energy band, and $\sigma_0$ is the normalization. We have found that
for the considerations in the present work, the cross section can be
safely approximated by the single pole of the Narrow-Width Approximation,
\begin{equation}
\sigma_A(\epsilon') = \pi\,\,\sigma_0\,\,  \frac{\Gamma}{2} \,\,
\delta(\epsilon' - \epsilon'_0)\, ,
\label{sigma}
\end{equation}
where $\sigma_0/A = 1.45\times 10^{-27} {\rm cm}^2$, $\Gamma = 8~{\rm
  MeV}$, and  $\epsilon'_0 = 42.65 A^{-0.21} \, (0.925 A^{2.433})~{\rm
  MeV},$ for $A > 4$ ($A\leq 4$)~\cite{KT93,noteSR}.
Inserting  Eq.~(\ref{sigma}) into Eq.~(\ref{rate}) we obtain
\begin{eqnarray}
R_A & \approx & \frac{\pi\, \sigma_0\,\epsilon'_0\, \Gamma}{4\,
  \gamma^2}
\int_0^\infty \frac{d \epsilon}{\epsilon^2}\,\,\, \nep \,\,\,
\Theta (2 \gamma \epsilon - \epsilon'_0) \nonumber \\
 & = & \frac{\pi \, \sigma_0 \,\epsilon'_0\, \Gamma}{4 \gamma^2}
\int_{\epsilon'_0/2 \gamma}^\infty \frac{d\epsilon}{\epsilon^2}\,\,
  \nep \,.
\label{kk}
\end{eqnarray}
As a test of our approximation, we first compute the disintegration
rate for a nucleus passing through a region where
\begin{equation}
\nep =  n^{\rm BE}_T (\epsilon) = (\epsilon/\pi)^2\
\left[e^{\epsilon/T}-1 \right]^{-1} \,\,,
\label{nBE}
\end{equation}
corresponding to a Bose-Einstein distribution with temperature $T$.
The result is
\begin{equation}
R_A^{{\rm BE}} \approx  \frac{\sigma_0 \,\epsilon'_0\,
\Gamma\,T}{4 \gamma^2 \pi} \,\,
| \ln \left(1 - e^{-\epsilon'_0/2 \gamma T}\right) | \,\,.
\label{RBE}
\end{equation}
We then verify that for $^{56}$Fe and for the CMB temperature ($T =
2.3 \times 10^{-4}~{\rm eV}$), this solution agrees to within 20\%
with the parametrization given in~\cite{Anchordoqui:1997rn}. The
latter was derived using the full form of the cross section presented
in~\cite{PSB76}. (A similar result in the context of
photo-disintegration of nuclei at the Galactic center has been
obtained in~\cite{Grasso:2005wd}.)

In the companion paper~\cite{shortP} it was argued that the rate can be
written as a function
$w^2\,|\ln (1-e^{-w})|$ of a scaling variable
$w\equiv \epsilon'_0/2\gamma T$, times a prefactor
$(\sigma_0\,\Gamma/\epsilon'_0)\,T^3/\pi$.
The peak rate therefore scales as the prefactor.
Using the fits mentioned below Eq.~(5), the prefactor in turn scales in
$A$ as $A^{1.21}$,
with a small correction for $A\le 4$ nuclei such as helium.

\subsection{The Photon Population}

The ingredients necessary for calculating the total photo-disintegration rate
from a given region of the OB association are {\it (i)} an ambient
photon distribution in order to obtain the rate $R_{A}^{\star}$ on
starlight per nucleon, and {\it (ii)} an initial population density of nuclei $n_A$
in the region. In a qualitative manner, the iron (silicon)
photo-disintegration rate in a region of radius $R_{\rm in} \approx 10~{\rm
  pc}$
\begin{eqnarray}
\frac{dN}{dt}& = &- N_{\rm Fe (Si)} \,\, R_{56 (28)}^{\star}
\nonumber\\
&=& - (4/3)\pi R_{\rm in}^3 \,\,\,n_{{\rm Fe (Si)}} \,\,\,R_{56(28)}^{\star} \,
,
\label{diegool}
\end{eqnarray}
where $n_{\rm Fe (Si)}$ is the iron (silicon) population density in
some energy bin. This population will be assumed to result from
continuous trapping of the diffuse cosmic ray flux by diffusion in a
milligauss magnetic field~\cite{Lai:2001sp}, acting preferably on
heavy nuclei.
Energies of ${\cal O}$ (PeV/nucleon) are achieved through
re-acceleration in strong winds of the OB stars.
We assume, and justify {\em a posteriori},
that the nucleus population $n_{\rm Fe (Si)}$ does not significantly change
during the time considered.
It will be seen below that {\em at most one nucleon will
be stripped in the region of interest during the diffusion time within
the association}; in the calculation of the disintegration rate,
this justifies inclusion of only the lowest order in the photo-nuclear
interaction.

The photon background will be assumed to result from the thermal
emission of the stars in the region $R$. The average density in
the region $R$ will reflect both the temperatures $T_{\rm O}$ and
$T_{\rm B}$ due to emission from O and B stars, respectively,  and the
dilution resulting from inverse square law
considerations. Specifically, for a region with $N_{\rm O}$ O stars
and $N_{\rm B}$ B stars, the photon density is
\begin{equation}
n^{\star}(\epsilon) = \frac{9}{4} \
\left[\frac{n^{\rm BE}_{T_{\rm O}}(\epsilon) \,\,
N_{\rm O} \,R_{\rm O}^2 + n^{\rm BE}_{T_{\rm B}}(\epsilon) \,\,
N_{\rm B} \,\,R_{\rm B}^2}{R^2} \right] \,\,,
\label{Jack}
\end{equation}
where $R_{\rm O(B)}$ is the radius of the O (B) stars, the factor 9/4
emerges when averaging the inverse square distance of an observer from
uniformly distributed sources in a region $R$~\cite{note94}. We take
the O and B star populations to be 5\% and 95\%, respectively, of the
total OB population of 2600 in the association. At this point, we do
not differentiate between the 3~pc cell HEGRA hot spot and the rest of
the cells in the association.  In order to take into account the
larger density of stars in the region of interest, we consider all the
stars to lie within the central region, $R = R_{\rm in} \sim 10$~pc.

\begin{figure}
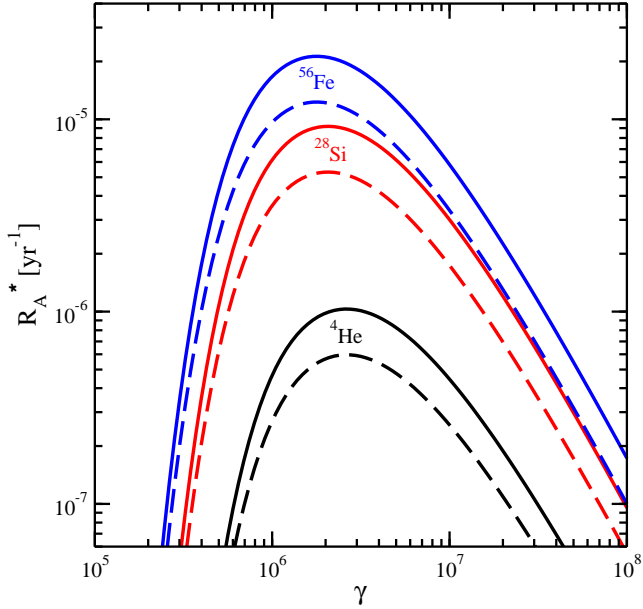

\begin{center}
\postscript{Rstars.eps}{0.98}
\caption{Photo-disintegration rate of $^{56}$Fe, $^{28}$Si, and $^4$He
  on the Cygnus OB2 starlight. Solid (dashed) lines represent the
  simplified (more elaborate) model as described in the text.}
\label{Rstar}
\end{center}
\end{figure}

In Fig.~\ref{Rstar} we show the dependence on the Lorentz factor of
$R_{56}^\star,$ $R_{28}^\star,$ and $R_4^\star$ for the stellar
ambiance described above (solid lines). The peaks of the three rate
curves, while occurring at slightly different values of $\gamma$, are
seen to scale as $A^{1.2}$.  This is in accord with our discussion of
scaling below Eq.~(\ref{RBE}).  Also in accord with our scaling
formula, the positions in $\gamma$ of the peak rates should scale in
$A$ as $\gamma_{\rm peak} \sim \epsilon'_0/w_{\rm peak}\sim A^{-0.21}$.
The three curves bear this out. We have taken for the O
(supergiants~\cite{supergiants}) stars a surface temperature $T_O$ =
40000~K, and radius $R_O = 19\ R_\odot$; for the cooler B stars we
assign $T_B$ = 18000~K and radius $R_B = 8\ R_\odot$~\cite{VGS96,H03}.
For all three nuclei, the disintegration time exceeds the diffusion time
($\sim 10^4~{\rm yr}$~\cite{B03}) of the nucleus in the association.
Thus, the {\em a priori} assumption of a lowest order calculation has
been justified~\cite{notexxx}.

At first sight this model might seem to be a very crude approximation
of the actual distribution of stars. We have also calculated the
photo-disintegration rate with a more detailed model of the star distribution.
We consider the thermal emission to come from the stars in the whole
association, $R_{\rm out}\sim$~30 pc, and that half of them are
uniformly distributed in the inner region, $R_{\rm in} \sim$~10 pc, and
the other half are uniformly distributed in the outer shell, i.~e. the
density of stars in the inner region is $(R_{\rm out}/R_{\rm in})^3 -
1 \sim 26$ times that in the outer shell. The photo-disintegration is
calculated to occur in a region of radius $r \sim $~3 pc in the
outskirts of the inner part of the association, $R_{\rm in}$, to model the
size and position of the source of the HEGRA signal.
This results in a photon density
\begin{equation}
n^{\star}(\epsilon) = \frac{47}{4} \ \left[\frac{n^{\rm
BE}_{T_{\rm O}}(\epsilon) \,\, N_{\rm O} \,R_{\rm O}^2 + n^{\rm
BE}_{T_{\rm B}}(\epsilon) \,\, N_{\rm B} \,\,R_{\rm B}^2}{R_{\rm
out}^2} \right] \,\,.
\end{equation}
The factor $47/4$ is a consequence of averaging the inverse
square distance within this distribution for the density and the
region where the reaction takes place~\cite{note94b}. In
Fig.~\ref{Rstar} we also show these results (dashed lines), which
agree very well with the previous less elaborate model for the
distribution of stars. This shows that even if we do not have a
quantitative explanation for the anisotropy at the border,
different global descriptions of the association point to the same
photo-disintegration rate, and the model we are proposing is stable.
It is clear, however, that within the 3~pc
HEGRA hot spot the
concentration of stars would be above average, and thus hereafter we
take as a fiducial value for $R^\star_A$ the one resulting from
Eq.~(\ref{Jack}) (more on this below).

After the high energy nuclei interact with the photon field entering
the GDR energy region, the nucleus is left in an excited state which
will go over into an excited daughter state emitting
$\gamma$-rays~\cite{BKM87}. Some early semiquantitative
statistical-model calculations for the production of $\gamma$-rays
through the decay of the GDR in the $^{56}$Fe nucleus showed that the
mean energy of the $\gamma$-spectrum is $\overline{E'_{\gamma 56}}
\sim 2-4$~MeV and the average multiplicity is $\overline{n_{56}} \sim
1-3$~\cite{Fe}. Previous measurements showed that for the case of
$^{16}$O, the corresponding values are $\overline{E'_{\gamma 16}} \sim
5-7$~MeV and $\overline{n_{16}} \sim 0.3-0.5$~\cite{CFB67}. Hence, in
the observer system, these relativistic nuclei are a source of
directional $\gamma$-rays with energy of the order $\sim \gamma$ MeV.

\subsection{TeV Gamma Ray Emission}

The low energy cutoff on $R_A^\star$ seen in Fig.~\ref{Rstar} will
be mirrored in the resulting photon distribution. The $E^{-2}$
energy behavior of the various nuclear fluxes will not
substantially affect this low energy feature.  It is a robust
consequence of the model, discussed in and below
Eq.~(\ref{qconst}). The energy behavior for photons in the
$1-10$~TeV region of the HEGRA data is a complex convolution of
the energy distributions of the various nuclei participating in
the photo-disintegration, with the rate factors appropriate to the
eV photon density for the various stellar populations.

Let us define $dR_A/dE'_\gamma$ as
\begin{equation}
\frac{dR_A}{dE'_\gamma} = \frac{1}{2} \, \int_0^{\infty}
\frac{n(\epsilon)}{\gamma^2 \epsilon^2}  \, d\epsilon \,
\int_0^{2\gamma \epsilon} \epsilon' \,
\frac{d\sigma_{\gamma A}}{dE'_\gamma}(\epsilon',E'_\gamma) \,
d\epsilon' \,\,,
\end{equation}
where $d\sigma_{\gamma A} (\epsilon',E'_\gamma)/dE'_\gamma$ is the
inclusive differential cross section for production of $\gamma$-rays
from disintegration and $E'_\gamma$ is the energy of the emitted
photon(s) in the rest frame of the nucleus. Assuming the same cosmic ray
spectrum as above, the emissivity (number/volume/steradian)
of $\gamma$-rays coming from nuclei
de-excitation can be written as
\begin{widetext}
\begin{equation}
Q_\gamma^{\rm dis} (E_\gamma)  =  \sum_A \int \frac{dn_A}{dE_N}(E_N) \,
dE_N \,\, \int  \frac{dR_A}{dE'_\gamma} \, \delta[E_\gamma - \gamma
  E'_\gamma (1+\cos{\theta_\gamma})] \, dE'_\gamma \,
\frac{d\cos{\theta_\gamma}}{2} \,\,,
\end{equation}
\end{widetext}
where $E_\gamma$ is the energy of the emitted $\gamma$-ray in the lab and
$\theta_\gamma$ is the $\gamma$-ray angle with respect to the
 direction of the excited nucleus.
We write the nuclear flux as
\begin{equation}
\sum_A \frac{dn_A}{dE_N}(E_N) = \sum_A N_A
\left(\frac{E_N}{E_0}\right)^{-\alpha} \,\,, \label{suma}
\end{equation}
with $N_A$ a normalization constant, and $E_0$ set to 1~TeV.
 Performing the angular integral
 with the delta-function constraint leads to
\begin{widetext}
\begin{equation}
Q_\gamma^{\rm dis} (E_\gamma)  = \sum_A \frac{m_N}{2} \int_{\frac{m_N
    E_\gamma}{2Q}} \frac{dn_A}{dE_N}(E_N) \, \frac{dE_N}{E_N}
  \int_{\frac{m_N E_\gamma}{2E_N}}^Q \frac{dR_A}{dE'_\gamma} \, \,
\frac{dE'_\gamma}{E'_\gamma} \,\,,
\end{equation}
\end{widetext}
where $Q$ is the $Q$-value of the de-excitation process. If we further
approximate the $\gamma$-ray spectrum as being monochromatic, with
energy equal to its average value ($\overline{E'_{\gamma A}}$), we can
write $d\sigma_{\gamma A}(\epsilon',E'_\gamma)/dE'_\gamma =
\overline{n_A} \, \sigma_A(\epsilon') \, \delta(E'_\gamma -
\overline{E'_{\gamma A}})$, where $\overline{n_A}$ is the mean
$\gamma$-ray multiplicity for a nucleus with atomic number $A$. Hence,
the emissivity can be approximated by~\cite{BKM87}
\begin{equation}
\label{qgdis}
Q_\gamma^{\rm dis}(E_\gamma) = \sum_A \frac{\overline{n_A} m_N}{2
  \overline{E'_{\gamma A}}} \int_{\frac{m_N E_{\gamma
  }}{2\overline{E'_{\gamma A}}}} \frac{dn_A}{dE_N}(E_N) \, R_A \,
  \frac{dE_N}{E_N} \, .
\end{equation}
The $\gamma$-ray emissivity is related to the differential flux at the
observer's site (assuming there is no absorption) as
\begin{equation}
\label{fluxg}
\frac{dF_\gamma}{dE_\gamma} (E_\gamma) =
\frac{V_{\rm dis}}{4 \pi d^2} \, Q_\gamma^{\rm dis} (E_\gamma)
\end{equation}
where $V_{\rm dis}$ is the volume of the source (disintegration) region
and $d$ is the distance to the observer.

It is clear from Eq.~(\ref{qgdis})
that if $R_A^\star$ is weakly dependent on $E_N$ then the
observed $\gamma$-ray flux will display the same power law behavior
as the nuclei population. The HEGRA data show an approximate
$E^{-2}$ behavior for $1\ {\rm TeV}\alt E_\gamma \alt 10\ {\rm TeV}$,
corresponding to a boost factor $10^6 \alt \gamma \alt 10^7$.
Remarkably, as can be seen in Fig.~\ref{Rstar}, $R_A^\star$ varies by
a factor of only 2 precisely in this region, far less than the 2
orders of magnitude of the primary nucleus flux. In order to evaluate
Eq.~(\ref{qgdis}) we approximate the behavior of $R_A^\star$ as
roughly constant for $E_{\rm min} < E_N < E_{\rm max}$, and zero
otherwise; where $E_{\rm min} \sim 10^6~{\rm GeV}$ and $ E_{\rm
  max} \sim 10^7~{\rm GeV}$.  Incorporating Eq.~(\ref{suma}) into
Eq.~(\ref{qgdis}), we find that  for $\alpha = 2$ and
$E_\gamma < \frac{2\overline{E'_{\gamma A}}\ E_{\rm
min}}{m_N}\simeq $ 2~TeV $(\overline{E'_{\gamma A}}/1\ {\rm MeV})$
\begin{equation}
Q_\gamma^{\rm dis}(E_\gamma) = \sum_A \frac{\overline{n_A} m_N}{4
  \overline{E'_{\gamma A}}}\ R_A^\star\ N_A \left(\frac{E_{\rm
  min}}{E_0}\right)^{-2} \,\,,
\label{qconst}
\end{equation}
{\it independent of} $E_\gamma$, whereas
for $E_\gamma > \frac{2\overline{E'_{\gamma A}}\ E_{\rm
min}}{m_N}\simeq $ 2~TeV~$(\overline{E'_{\gamma A}}/1\ {\rm MeV})$
\begin{equation}
Q_\gamma^{\rm dis}(E_\gamma) = \sum_A \frac{\overline{n_A} m_N}
{4 \overline{E'_{\gamma A}}}\ R_A^\star\ \ N_A\
\left(\frac{E_\gamma}{E_{\gamma 0}}\right)^{-2}\ \ ,
\label{qvary}
\end{equation}
where
\begin{eqnarray}
E_{\gamma 0} &=& \frac{2 \overline{E'_{\gamma A}}E_0}{m_N} \nonumber
   \\[.1in]
   &\simeq & 2000 \overline{E'_{\gamma A}}\ \ .
\end{eqnarray}
The predicted constancy of the flux below $\sim$ 2 TeV implies a
strong suppression in this region relative to a flux extrapolated
from HEGRA data to maintain an $E_\gamma^{-2}$ behavior down to
lower energies. For example, the upcoming GLAST mission will probe
the gamma ray spectrum from the Cygnus region in the range
$20~{\rm MeV} - 300~{\rm GeV}$~\cite{Carson:2006af}. The flux
predicted at $\sim$ 100 GeV from an $E_\gamma^{-2}$ extrapolation
of the HEGRA data would render the source spectacularly visible in
the GLAST observation, whereas the model here would predict a
suppression by a factor of $\sim\ (2/0.1)^2 = 400$ relative to
this extrapolated flux. At 500 GeV the suppression would be still
statistically significant: $E_\gamma^{-2} \times$ flux falls by a
factor of $\sim 16.$

\begin{figure}
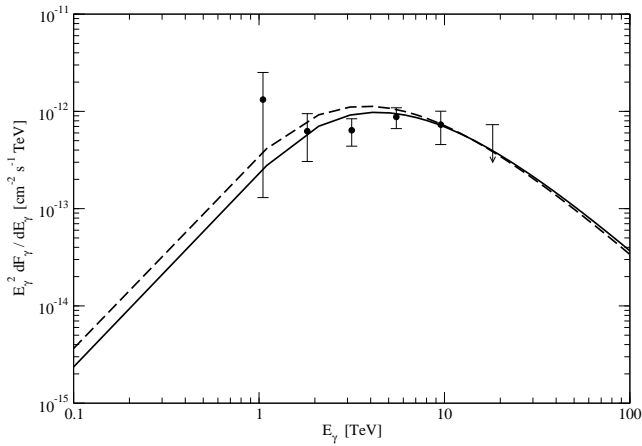

\begin{center}
\postscript{HEGRA_LPE2.eps}{0.98}
\end{center}
\caption{Energy-weighted $\gamma$-ray flux superposed over the HEGRA data.
  The curves are not actual fits to the spectrum but are based on
  particular choices of the parameters which provide eyeball agreement
  with the data. The solid line corresponds to $^{28}$Si and the
  dashed one to $^{56}$Fe.  We have used $N_{28} = 2 \times 10^{-10} {\rm
    cm}^{-3} {\rm TeV}^{-1}$ and $N_{56} = 10^{-10} {\rm cm}^{-3} {\rm
    TeV}^{-1}.$ In both cases, $\overline{E_{\gamma A}'}$ = 1.5~MeV
  and the spectral index $\alpha=2$. Note that the choice
  $\overline{E_{\gamma A}'}$ = 1.5~MeV differs from the value 1~MeV
  which was used as fiducial input for the discussion in the text.}
\label{specs}
\end{figure}

The previous discussion has been qualitative.  In Fig.~\ref{specs}
these features are display following a direct integration of
Eq.~(\ref{qgdis}), with choice of parameters to provide eyeball
agreement with the data. One can see that the behavior of the fully
integrated spectrum agrees in essential characteristics with the
results of the preceding qualitative discussion.

All in all, in the present model the photon flux
follows the parent population of nuclei in the PeV/nucleon energy
region. Consistency then requires the nucleus spectrum not to be
significantly steeper than $E^{-2}$. How does this comply with
experimental data?

As mentioned in Sec.~\ref{II}, the absence of supernova remnants in the
association suggests that the nucleus population originates in the
diffuse cosmic ray flux rather than in local supernova explosions.
In assessing the consistency of the model we first obtain the
nuclear population with $\gamma\sim 10^6$ (i.e., PeV~energy/nucleon)
necessary for matching the HEGRA data.
For simplicity, for this purpose, we approximate the sum in
Eq.~(\ref{qgdis}) with a single species (take silicon). Setting the
volume $V=(4/3)\pi r^3$ and the spectral index $\alpha =2,$ we find
from Eqs.~(\ref{Hflux}) and (\ref{fluxg}) the normalization constant
\begin{equation}
N_{28} = 1.0 \times 10^{-10}\ {\rm cm}^{-3}\ {\rm TeV}^{-1}\ \ .
\label{n0}
\end{equation}
From Eq.~(\ref{suma}) this gives a required Si density at 1~PeV
\begin{eqnarray}
\frac{dn_{\rm Si}}{dE_{\rm Si}} (E_N = 1 \, {\rm PeV}) & = &
\frac{1}{28}\ \frac{dn_{\rm Si}}{dE_N} (E_N = 1 \, {\rm PeV})
\nonumber \\
 & \approx & 3.3\times 10^{-18}\ {\rm cm}^{-3}\ {\rm TeV}^{-1}\ \ .
\end{eqnarray}
It is of interest to compare this with the observed diffuse cosmic ray
nuclear density $dn_{\rm CR,Si}/dE$ in this energy region.
From~\cite{Horandel:2005at}, we find
\begin{eqnarray}
\frac{dn_{\rm CR,Si}}{dE} (1 \, {\rm PeV}) & = & \frac{4\pi}{c} \, \,
J_{\rm CR,Si}(1~{\rm PeV}) \nonumber \\
 & \approx & 1.3\times 10^{-23}\ {\rm cm}^{-3}\ {\rm TeV}^{-1},
\label{JCR}
\end{eqnarray}
a difference of 5 orders of magnitude. Combined with the
observation that the diffuse spectrum has a power index of 3, this
shows explicitly that {\it (i)} the required population density at
PeV/nucleon must arise from acceleration of much more plentiful
nuclei that are trapped at much lower energies {\it (ii)} the
acceleration efficiency of the association must be considerably
greater than that of the Galaxy in order to provide a harder
spectrum $\propto E^{-2}.$ In order to ascertain whether this is
possible, we compare the total energy in the nucleus population
with the wind energy budget $\sim 10^{39}$~erg/s. Integration of
$E_N dn_{\rm Si}(E_N)/dE_N$, with the use of Eq.~(\ref{n0}), gives
for the volume of the HEGRA source a total energy $\sim
10^{48}$~erg (modulo a logarithmic dependence on the source
cutoff). This energy is accumulated over the diffusion time of
$10^4\,{\rm yr}=3\times 10^{11}$~seconds, so that the required
power density is 2 orders of magnitude smaller than the kinetic
energy budget of the entire association.

In summary, we have presented in detail the criteria for the model to
be viable.  If future observations should lend credence to the model,
significant insights into the acceleration mechanism can be obtained,
especially to provide clues with respect to the acceleration
efficiency. In completing the explanation of the HEGRA signal, there
is of course one issue that needs to be addressed -- the signal was
observed only in a 3 pc radius cell at the edge of the association.
With the same angular radius, there are a total of 37 cells in the
core of the association. The flux in each of the other cells is
limited to be $\alt 0.17$ of TeV J2032+4130. At this point in our
understanding we can provide only qualitative remarks. The obvious
possibility is an increased density of very hot OB stars in the cell
of TeV J2032+4130 which provide more efficient trapping and
accelerating conditions, as well as a hotter photon background.
Indeed, a recent estimate~\cite{Butt03} indicates around 10 O stars in
the region of the source, which is a factor of 3 larger than expected
on the basis of a uniform population. (This strongly supports our
previous choice of $R^\star_A$.) More understanding of this can only
come with more data and a full Monte Carlo simulation.

\subsection{TeV Neutrino Emission}

As mentioned above, the interaction of nuclei with the background
photons produces a beam of neutrons.  The decay mean free path of a
neutron is $c\,\gamma \,\overline\tau_n=9.15\,(E_n/10^9~{\rm
  GeV})$~kpc, the lifetime being boosted from its rest-frame value
$\overline\tau_n=886$~seconds to its lab value via $\gamma =E_n/m_n$.
This means that for a source distance $d \sim 1.7~{\rm kpc},$ practically
all neutrons with $E_n \sim 10^6~{\rm GeV}$ will decay {\em en route}
to Earth, producing a flux of directional antineutrinos.

The basic formula that relates the neutron flux at the source
($d\Fn/d\En$) to the antineutrino flux observed at Earth
($d\Fnu/d\Enu$) is~\cite{Anchordoqui:2003vc}
\begin{widetext}
\begin{equation}
\frac{d\Fnu}{d\Enu}(\Enu)  =
\int d\En\,\frac{d\Fn}{d\En}(\En)
\left(1-e^{-\frac{m_n\,d}{\En\,\tbar}}\right)\,
\int_0^{Q_{\overline \nu}} d\Enubar\,\frac{dP}{d\Enubar}(\Enubar)
\int_{-1}^1 \frac{d\cth}{2}
\;\delta\left[\Enu-\En\,\Enubar\,(1+\cth)/m_n\right]
\,.
\label{nuflux}
\end{equation}
\end{widetext}
The variables appearing in Eq.~(\ref{nuflux}) are the antineutrino and
neutron energies in the lab ($\Enu$ and $\En$), the antineutrino angle
with respect to the direction of the neutron momentum, in the neutron
rest-frame ($\thnu$), and the antineutrino energy in the neutron
rest-frame ($\Enubar$).  The last three variables are not observed by
a laboratory neutrino-detector, and so are integrated over.  The
observable $\Enu$ is held fixed.  The delta-function relates the
neutrino energy in the lab to the three integration variables. The
parameters appearing in Eq.~(\ref{nuflux}) are the neutron mass and
rest-frame lifetime ($m_n$ and $\tbar$). Finally, $dP/d\Enubar$ is the
normalized probability that the decaying neutron produces a $\overline
\nu$ with energy $\Enubar$ in the neutron rest-frame. Note that the
maximum $\overline \nu$ energy in the neutron rest frame is very
nearly $Q_{\overline \nu} \equiv m_n - m_p - m_e = 0.78$~MeV.
Integration of Eq.~(\ref{nuflux}) can be easily accomplished,
especially when two good approximations are applied. In the lab, the
ratio of the maximum $\overline \nu$ energy to the neutron energy is
$Q_{\overline \nu}/m_n \sim 10^{-3},$ and so the boosted $\overline
\nu$ has a spectrum with $E_{\overline\nu} \in (0, 10^{-3} \,
E_n)$ and an average energy $\epsilon_0 = 0.48$~MeV. The first
approximation is to simplify the antineutrino spectrum from
$\beta$-decay and take a monochromatic spectrum with energy equal to
the average energy, $\epsilon_0$. In such a way $dP/d\Enubar =
\delta(\Enubar - \epsilon_0)$ and Eq.~(\ref{nuflux}) simplifies. The
second approximation is to replace the neutron decay probability $1 -
e^{-d m_n/E_n \overline \tau_n}$ with a step function $\Theta
(E_n^{\rm max} - E_n)$ at some energy $E_n^{\rm max} \sim {\cal
  O}(m_n\,d /\overline{\tau}_n) = (d/9.15~{\rm kpc}) \times
10^{9}$~GeV. Combining these two approximations,
one obtains~\cite{Anchordoqui:2003vc}
\begin{equation}
\label{nuflux4}
\frac{d\Fnu}{d\Enu}(\Enu)=\frac{m_n}{2\,\eps_0}
\int^{\Enmax}_{\frac{m_n\,\Enu}{2\,\eps_0}} \frac{d\En}{\En}\,
    \frac{d\Fn}{d\En}(\En)\,.
\end{equation}

The neutron emission can be related to the HEGRA $\gamma$-ray flux
only if photo-de-excitation following photo-dissociation is
unsuppressed. Thus, a {\em lower} bound for neutron emissivity
(and resulting neutrino flux) may be obtained by finding this
relation for the cases (Si/Fe) discussed in the previous section.
An extremely important example where there is 100\% suppression is
the case of $^4$He, because neutron emission results in
transitions to stable $A=3$ daughter states~\cite{nuclides}.
This possibility will be discussed more at the end of this section.
Meanwhile, we turn
to estimate the neutrino event rate associated with Si dissociation which
can be expected at the IceCube detector, now under construction at
the South Pole~\cite{Ahrens:2003ix}.

For a transition in which an average of $\overline {n_{28}}$
$\gamma$-rays are emitted during de-excitation, an average of $\sim$ 1/2
neutron is emitted during the stripping (there is almost equal
probability of emission of $n$ and $p$). The conservation of Lorentz
factor allows the relation
\begin{equation}
\int^{\gamma  \overline{E'_{\gamma 28}}} \frac{dF_\gamma}{dE_\gamma}
\,\,dE_\gamma = 2\, \overline{n_{28}}\,\, \int^{\gamma  m_n}
\frac{dF_n}{dE_n} \, dE_n \, .
\end{equation}
Taking the derivative with respect to $\gamma$ gives the relation
\begin{equation}
\overline{E'_{\gamma 28}} \left. \,\,\frac{dF_\gamma}{dE_\gamma}
\right|_{E_\gamma = \gamma \overline{E'_{\gamma 28}}} =
2 \,m_n \,\overline{n_{28}}\,\, \left.
\frac{dF_n}{dE_n}\right|_{E_n = \gamma m_n} \,,
\end{equation}
which leads to desired relation between photon and neutron fluxes
\begin{equation}
\frac{dF_n}{dE_n} (E_n) = \frac{\overline{E'_{\gamma 28}}}
{2 m_n \overline{n_{28}}}\,
\,\,\,\frac{dF_\gamma}{dE_\gamma}
(E_\gamma = E_n \overline{E'_{\gamma 28}}/m_n) \, .
\end{equation}
Substituting into Eq.~(\ref{nuflux4}), one finds the antineutrino flux
associated to a given flux of photons
\begin{equation}
\label{qnudis}
\frac{d\Fnu}{d\Enu}(\Enu) = \frac{1}{\alpha}
\left(\frac{2 \epsilon_0}{\overline{E'_{\gamma 28}}}\right)^{\alpha
  -1} \, \frac{1}{2 \overline{n_{28}}}
\,\,\,\frac{dF_\gamma}{dE_\gamma} (E_\gamma = \Enu)\, ,
\end{equation}
where $\alpha$ is the spectral index of the photon population. When
referring to the 3 pc HEGRA cell, and taking $2\epsilon_0 \simeq
\overline{E'_{\gamma 28}}$, we finally obtain the predicted associated
antineutrino flux
\begin{equation}
\frac{d\Fnu}{d\Enu}(\Enu) = \frac{1}{4} \,\frac{N_{\rm HEGRA}
}{ \overline{n_{28}}}  \,\, \left(\frac{E_{\overline \nu}}{\rm TeV}
\right)^{-\alpha_{_{\rm HEGRA}}} \,\,.
\end{equation}
which is valid in the energy window for which nuclei disintegration
takes place, i.e., $\Enu \sim (10^6 - 10^7) \, \epsilon_0$.

At IceCube, the events are grouped as either muon tracks or
showers. Tracks include muons resulting from both cosmic muons and
from Charged Current (CC) interaction of muon neutrinos. The angular
resolution for muon tracks $\approx 0.7^\circ$~\cite{Ahrens:2002dv}
allows a search window of $1^\circ \times 1^\circ$. This corresponds
to a search bin solid angle of $\Delta \Omega_{1^\circ \times 1^\circ}
\approx 3 \times 10^{-4}$~sr. Since IceCube does not resolve the 3~pc
HEGRA cell, the $\overline \nu$ flux will have contribution from all
37 cells in the core of the association. To estimated this, we assume
that the flux from each of the 36 non-HEGRA cells is equal to the upper
limit found in the direction of Cygnus X-3, namely 1/6 of the flux
from TeV J2032+4130. The antineutrino flux from the Cygnus region  can
then be approximated (perhaps generously) as
\begin{equation}
\frac{d\Fnu}{d\Enu}(\Enu) = \frac{7}{4} \,\frac{N_{\rm HEGRA}
}{ \overline{n_{28}}}  \,\, \left(\frac{E_{\overline \nu}}{\rm TeV}
\right)^{-\alpha_{_{\rm HEGRA}}} \,\,.
\label{antinus}
\end{equation}

To estimate the expected number of $\overline{\nu}_\mu$ induced tracks
from Cygnus OB2 we adopt the semianalytical calculation presented
in Ref.~\cite{Gonzalez-Garcia:2005xw},
\begin{widetext}
\begin{equation}
{\cal N}_{\rm Cyg}^{\rm tr}
 = t \, n_{\rm T}\,
\int^\infty_{l'_{\rm min}}  dl\,
\int_{m_\mu}^\infty  dE_\mu^{\rm fin}\,
\int_{E_\mu^{\rm fin}}^\infty  dE_\mu^0\,
\int_{E_\mu^0}^\infty  dE_{\overline \nu}
\frac{ d F_{\overline \nu_\mu}}{ dE_{\overline \nu}}(E_{\overline \nu})
\frac{d\sigma_{\rm CC}}{ dE_\mu^0}(E_{\overline \nu},E_\mu^0)\,
F(E^0_\mu, E_\mu^{\rm fin}, l)\, A^0_{\rm eff}\, \,,
\label{eq:trsource}
\end{equation}
\end{widetext}
where $d F_{\overline \nu_\mu}/dE_{\overline \nu}$ is the $\overline
\nu_\mu$ flux, $d\sigma_{\rm CC}/E_\mu^0$ is the differential CC
interaction cross section producing a muon of energy $E_\mu^0$,
$n_{\rm T}$ is the number density of nucleons in the matter
surrounding the detector, and $t$ is the exposure time of the
detector. After being produced, the muon propagates through the rock
and ice surrounding the detector and loses energy. We denote by
$F(E^0_\mu,E_\mu^{\rm fin},l)$ the function that represents the
probability of a muon produced with energy $E_\mu^0$, arriving at the
detector with energy $E_\mu^{\rm fin}$, after traveling a distance
$l$. The details of the detector are encoded in the effective area
$A^0_{\rm eff}$. We use the parametrization of the $A^0_{\rm eff}$
described in Ref.~\cite{Gonzalez-Garcia:2005xw} to simulate the
response of the IceCube detector after events that are not neutrinos
have been rejected (this is achieved by quality cuts referred to as
``level 2'' cuts in Ref.~\cite{Ahrens:2003ix}). The minimum track
length cut is $l_{\rm min}=300$~m and we account for events with
$E_\mu^{\rm fin}>500$~GeV.

Although the flux of antineutrinos produced by Cygnus OB2 is pure
$\overline \nu_e,$ because of neutrino oscillations, the antineutrinos
observed at Earth will be distributed over all flavors,
\begin{equation}
\frac{ d F_{\overline\nu_\alpha}}{ dE_{\overline \nu}}
= \left(\frac{1}{3} +
f_{\overline\nu_e\to\overline\nu_\alpha}(E_{\overline \nu}) \right) \,
\frac{{\rm d} F_{\overline \nu}}{{\rm d}E_{\overline \nu}}\, ,
\label{osc}
\end{equation}
where $\alpha = e,\ \mu,\ \tau$ denotes the neutrino flavor. The 95\%
confidence ranges for the probability differences,
\begin{eqnarray}
f_{\overline \nu_e \to  \overline\nu_\mu}&=&-0.106^{+0.060}_{-0.082}
\,\,,\nonumber\\
f_{\overline \nu_e \to  \overline \nu_\tau}&=&-0.128^{+0.089}_{-0.055}
\,\,, \label{fit} \\
f_{\overline\nu_e \to \overline \nu_e} & = & - (f_{\overline \nu_e
  \to \overline\nu_\mu} + f_{\overline \nu_e \to  \overline
  \nu_\tau})\,, \nonumber
\end{eqnarray}
were derived elsewhere~\cite{Anchordoqui:2005gj} using the results of
the up-to-date 3$\nu$ oscillation analysis of solar, atmospheric, LBL
and reactor data~\cite{globalfit}.  Substituting Eqs.~(\ref{osc}) and
(\ref{fit}) into Eq.~(\ref{eq:trsource}) we obtain the
$\overline\nu_\mu$-induced tracks in IceCube from Cygnus OB2 in $t =
15$ years of observation, ${\cal N}^{\rm tr}_{\rm Cyg} \approx
7.5/\overline{n_{28}}.$

Showers are generated by neutrino collisions --- $\nu_e\ \mbox{or}\,\,
\nu_\tau$ CC interactions, and all Neutral Current (NC) interactions
--- inside of or nearby the detector, and by muon bremsstrahlung
radiation near the detector. For showers, the angular resolution is
significantly worse than for muon tracks. In our analysis, we consider
a shower search bin solid angle, $\Delta \Omega_{10^\circ \times
  10^\circ}.$ Normally, a reduction of the muon bremsstrahlung
background is effected by placing a cut of 40 TeV on the minimum
reconstructed energy~\cite{Ackermann:2004zw}. For Cygnus OB2, this
strong energy cut is not needed since this muon background is filtered by
the Earth. Thus we account for all events with shower energy $E_{\rm
  sh}\geq E_{\rm sh}^{\rm min}=1$~TeV. The directionality requirement,
however, implies that the effective volume for detection of showers is
reduced to the instrumented volume of the detector, ${\cal V}_{\rm
  eff} = 1~{\rm km}^3$, because of the small size of the showers (less
than 200 m in radius) in this energy
range. Following Ref.~\cite{Anchordoqui:2005gj} we estimate the expected
number of showers from Cygnus OB2 as
\begin{equation}
{\cal N}_{\rm Cyg}^{\rm sh}  =
{\cal N}_{\rm Cyg}^{\rm sh,CC}+{\cal N}_{\rm Cyg}^{\rm sh,NC} \,\,,
\end{equation}
where
\begin{widetext}
\begin{equation}
{\cal N}_{\rm Cyg}^{\rm sh,CC} =  t\, n_{\rm T}\, {\cal V}_{\rm eff}\,
\int_{E^{\rm min}_{\rm sh}}^\infty
{\rm d}E_{\overline \nu} \, \sum_{\alpha=e,\tau}\frac
{d F_{\overline \nu_\alpha}}{dE_{\overline \nu}}(E_{\overline \nu})
\sigma_{\rm CC}(E_{\overline \nu}) \,\,,
\label{eq:shsourcecc}
\end{equation}
and
\begin{equation}
{\cal N}_{\rm Cyg}^{\rm sh,NC} =  t\, n_{\rm T}\, {\cal V}_{\rm eff}\,
\int_{E_{\overline\nu}-E^{\rm min}_{\rm sh}}^\infty  dE'_{\overline
  \nu} \int_{E^{\rm min}_{\rm sh}}^\infty dE_{\overline \nu} \,
\sum_{\alpha=e,\mu,\tau}
\frac{ d F_{\overline \nu_\alpha}}{ dE_{\overline \nu}}(E_{\overline
  \nu}) \frac{d\sigma_{\rm NC}}{dE'_{\overline\nu}}
(E_{\overline \nu}, E'_{\overline \nu}) \,\,.
\label{eq:shsourcenc}
\end{equation}
\end{widetext}
Here, $d\sigma_{\rm NC}/dE'_{\overline\nu}$ is the differential NC
interaction cross section producing a secondary antineutrino of
energy, $E'_{\overline\nu}$. In writing Eqs.~({\ref{eq:shsourcecc}})
and (\ref{eq:shsourcenc}) we are assuming that for contained events
the shower energy corresponds with the interacting $\overline\nu_e$ or
$\overline\nu_\tau$ antineutrino energy ($E_{\rm
sh}=E_{\overline\nu}$) in a CC interaction, while for NC the shower
energy corresponds to the energy in the hadronic shower $E_{\rm
sh}=E_{\overline\nu}-E'_{\overline\nu}\equiv y\, E_{\overline\nu}\,$
where $y$ is the usual inelasticity parameter in deep inelastic
scattering. In total we expect ${\cal N}_{\rm Cyg}^{\rm sh} \approx
5/\overline{n_{28}}$ from Cygnus OB in 15 years of observation.

\begin{table*}
\begin{tabular}{|c|c|c|c|c|c|} \hline
Primary quanta & $\gamma$ production mech. & $\nu$ production mech. &
   Number ratio & Energy ratio & comments        \\ \hline
A & $A \rightarrow A^* \rightarrow \,\sim \overline{n_A} \, \gamma+A$ &
   $A \rightarrow n \rightarrow {\overline \nu}_e$ &  $\gamma:\nu \sim
   \overline{n_A}:1/2$ & $\frac{E_\gamma}{E_\nu} \sim
   \frac{\overline{E'_{\gamma A}}}{\epsilon_0} $ &  observation of $\nu$'s
   at IceCube \\
 & $[ E_\gamma \sim \gamma_{A} \, \overline{E'_{\gamma A}} ]$ & $[ E_\nu \sim
 \gamma_{A} \, \epsilon_0 ]$ & & \hspace{6mm} $\sim 2 - 8$ & depends on
$^4$He abundance\\ \hline
p & $p \rightarrow \pi^0 \rightarrow 2\,\gamma$ & $p\rightarrow
   \pi^{\pm}\rightarrow 3\,\nu$ & $\gamma:\nu \sim 1:3$ &
   $\frac{E_\gamma}{E_\nu}\sim 2$ & $\nu$'s ARE seen at IceCube\\
 & $[ E_\gamma\sim \frac{1}{10}\,E_p ]$ & $[ E_\nu\sim
     \frac{1}{20}\,E_p ]$ & & & \\ \hline $e$-plasma & synchrotron &
     none & & & \\
 & inverse Compton & & & & \\ \hline
\end{tabular}
\caption
{\label{table:comparison}
Comparison of $\gamma$-ray and neutrino emission from $A$, $p$, and $e$
primaries. Note that per $\gamma$-ray, an order of magnitude fewer
neutrinos are expected from nuclei photodisentigration than from
hadronic interactions followed by pion decays. Note also that the
neutrino energy from the nuclei photo-disintegration is typically about
one order of magnitude smaller than the $\gamma$-ray energy. When the
primaries are electrons, only $\gamma$-rays are produced, but not
neutrinos.}
\end{table*}

We now turn to the estimate of the background of atmospheric
neutrinos. For the ``conventional'' atmospheric neutrino fluxes
arising from pion and kaon decays, we adopt the 3-dimensional scheme
estimates of Ref.~\cite{Honda:2004yz}, which we extrapolate to match
at higher energies the 1-dimensional calculations by
Volkova~\cite{Volkova:1980sw}. We also incorporate ``prompt''
neutrinos from charm decay as calculated in
Ref.~\cite{Gondolo:1995fq}. We obtain the number of expected track and
shower events from atmospheric neutrinos as in
Eqs~(\ref{eq:trsource}), (\ref{eq:shsourcecc}), and
(\ref{eq:shsourcenc}) with $dF^{\rm
  ATM}_{\nu_\alpha}/dE_{\nu}(E_{\nu})$ being the $\nu_e$ and
$\nu_\mu$ atmospheric neutrino fluxes integrated over a solid angle
of $1^\circ\times 1^\circ$ (for tracks) and $10^\circ\times 10^\circ$
(for showers) width around the direction of the Cygnus OB2 source
$\theta=131.2^\circ$. We get an expected background of ${\cal N}^{\rm
  tr}_{\rm ATM} = 14$ and ${\cal N}^{\rm sh}_{\rm   ATM} = 47$ in 15
years. Of the 47 showers, 16 correspond to $\nu_e$ CC interactions
while 31 correspond to $\nu_\mu$ NC interactions.

It is clear from the preceding that there is no significant
antineutrino signal resulting from dissociations which are
accompanied by unsuppressed photo-de-excitation. This is in
complete contrast with the $pp$ case where for an spectral index
of 2 one expects on average a flux of $\nu_\mu$ half of that of
$\gamma$-rays~\cite{KHSA06,Gaisser,CV05,Anchordoqui:2004eu,Kistler:2006hp}
(see Sec.~\ref{V}). Then, when considering the 37 cells of the
Cygnus OB2 association one expects a $\nu_\mu$ event rate of
4.2~yr$^{-1}$ with a background of 0.9~yr$^{-1}$. The track
signal in the $pp$ case is about an order of magnitude larger
than in the $A^*$ case.  Because of production and oscillations
the $pp$ mechanism yields about one $\nu_\mu$ per $\gamma$-ray. The
yield in the $A^*$ mechanism (for $\overline{n_{28}} = 2$)
is about 0.25 $\overline \nu_e$ per
$\gamma$-ray which, after oscillation, yields 0.05 $\overline
\nu_\mu$ per $\gamma$-ray. This ratio decreases by an additional factor of
$\sim 2$ if comparison is made with integrated spectrum above 1 TeV.

We turn now to comment on the role of helium. Except for protons,
helium nuclei dominate the cosmic ray spectrum, with a population
about 100 times larger than the heavy
species~\cite{Yao:2006px,factor100}. As mentioned above,
the stripping of a nucleon from $^4$He leaves the residual $A=3$
nuclides in their ground states~\cite{nuclides}, so that there is
no photon emission. However, stripping to $^3$He with emission of
a neutron will provide a yield of neutrinos. As can be seen from
Fig.~\ref{Rstar}, the stripping rate is down by an order of
magnitude from that for the heavy elements so that, with the
larger population, the expected antineutrino flux would be about a
factor of 10 larger than our prediction from Eq.~(\ref{antinus}).
This enhancement has powerful consequences for the source
energetics: the energy required to accelerate the entire $^4$He
population is very close to the allowed energy budget for the
diffusion lifetime. Therefore, should IceCube obtain a
statistically significant signal that cannot be ascribed to $pp$
interaction (because of observation of the TeV $\gamma$-ray suppression
predicted by the model), then there could be a hint of
extraordinary efficiency in the trapping an accelerating
mechanisms in extremely hot and intense stellar associations.

\section{Photo-Disintegration \lowercase{vs} $\pi$ decay}
\label{V}

In the previous section we have described the mechanism of $\gamma$-ray
production from nuclei de-excitation after disintegration in the
background photon field and showed that the HEGRA data could be
explained in these terms. In addition, this mechanism can give rise
to a neutrino flux after the stripped neutrons decay in
flight. However, as we mentioned in the Introduction, there are
two channels other than photo-disintegration that might contribute to
$\gamma$-ray and neutrino production at the Cygnus region. These are
photo-hadronic ($A$-$\gamma$) and pure hadronic ($A$-$p$)
interactions. In both cases, $\gamma$-rays (neutrinos) are produced
after $\pi^0$ ($\pi^+$ and $\pi^-$) decays.

As noted above, photo-meson production has a very high energy
threshold, being only relevant for very high energetic beams or in
very hot photon enviroments. Even in these extreme cases, the fact
that this reaction turns on at so high energies implies that the
photons and neutrinos from decaying pions are produced at very
high energies too, well above the TeV range. Hence, in the following
subsections we comment on the $\gamma$-ray and neutrino emissivities
due to nucleus-proton collisions and compare it with those calculated
for the photo-disintegration of nuclei. But first, in
Table~\ref{table:comparison} we show approximate estimates for the
energy, number ratio and energy ratio of neutrinos and $\gamma$-rays
due to photo-disintegration and pion decay at production (we have also
included the leptonic mechanism of $\gamma$-ray production for
completeness). Now, we will compare these two hadronic mechanisms.

\subsection{$\pi$ spectrum}

The interaction of high energy nuclei with the cold ambient
interstellar medium (ISM) gives rise to $\gamma$-rays through the decay
of the produced neutral mesons. The $\pi^0$ emissivity resulting from
an isotropic distribution of accelerated nuclei $dn(E_N)/dE_N $ is
given by~\cite{S71}
\begin{eqnarray}
\label{Qpi}
Q_{\pi^0}^{Ap} (E_{\pi^0}) & = &  c \, n_{\rm H} \, \int_{E_N^{th}
  (E_{\pi^0})}^{E_N^{max}} \frac{dn}{dE_N} (E_N) \nonumber \\
 & \times &
  \frac{d\sigma_A}{dE_{\pi^0}} (E_{\pi^0},E_N) \, dE_N
\end{eqnarray}
where $n_{\rm H}$ is the ISM number density, $E_N^{th} (E_{\pi^0})$ is
the minimum energy per nucleon required to produced a pion with energy
$E_{\pi^0}$, and $d\sigma_A(E_{\pi^0},E_N)/dE_{\pi^0}$ is the
differential cross section for the production of a pion with energy
$E_{\pi^0}$ in the lab frame due to the collision of a nuclei $A$ of
energy per nucleon $E_N$ with a hydrogen atom at rest.

Hence, an accurate knowledge of the differential cross section for
pion production is necessary to calculate the $\gamma$-ray emissivity
from this channel. There have been several approaches and
parameterizations in the literature, most of them based on the use of
the isobaric~\cite{S70} and scaling~\cite{SB,BSKNN00} models of the
reaction or their combination~\cite{D86,SM98,BC99}. The
$\delta$-function approximation was considered in Ref.~\cite{AA00} and
the inclusion of diffractive interactions and scaling violations in
Ref.~\cite{ppdifscal} (see the Appendix of Ref.~\cite{DT0506} for a a
comparison of different approaches). A new recent parametrization is
presented in Ref.~\cite{KAB06} based on simulations of proton-proton
interactions from the SIBYLL event generator~\cite{SIBYLL}. The
isobaric model has been shown to work reasonably well at low energies
($E < 3$ GeV), whereas the scaling model is more suitable at higher
energies. Hence, here we will follow the scaling model with a
parameterization of the differential cross section which is an
$E_N$-independent approximation of that given in Ref.~\cite{KAB06}
\begin{equation}
\frac{d\sigma_A}{dE_{\pi^0}} (E_{\pi^0},E_N) \simeq
\frac{\sigma_0^A}{E_{\pi^0}} \, f_{\pi^0} (x)
\end{equation}
where $x \equiv E_{\pi^0}/E_N$ and $\sigma_0^A = A^{3/4} \, \sigma_0$,
with $\sigma_0 = 34.6$ mb, which takes into account the scaling of
the cross section with the atomic number~\cite{LST63}, and
\begin{widetext}
\begin{equation}
f_{\pi^0} (x) \simeq 8.18 \, x^{1/2} \,
  \left(\frac{1-x^{1/2}}{1+1.33 \, x^{1/2} \, (1-x^{1/2})}\right)^4 \,
  \left(\frac{1}{1-x^{1/2}} + \frac{1.33 \, (1-2 x^{1/2})}{1 + 1.33 \,
  x^{1/2} \, (1 - x^{1/2})}\right)
\end{equation}
\end{widetext}
which takes into account the high pion multiplicities at high
energies.

By using this form for the differential cross section and a power-law
cosmic-ray spectrum, the $\pi^0$ emissivity can be written as
\begin{equation}
Q_{\pi^0}^{Ap} (E_{\pi^0}) \simeq  Z_{A \pi^0} (\alpha) \, Q_A^{Ap}
(E_\pi^0)
\end{equation}
where
\begin{equation}
Q_A^{Ap} (E_N) = \sigma_0^A \, c \, n_{\rm H} \, \frac{dn}{dE_N} (E_N)
\end{equation}
and the spectrum-weighted moment of the inclusive cross section or
so-called $Z$-factor is given by
\begin{equation}
Z_{A \pi^0} (\alpha) \equiv \int_0^1 x^{\alpha-2} \, f_{\pi^0} (x) \,
dx
\end{equation}
where, as usual, $\alpha$ is the spectral index of the cosmic-ray
spectrum.

\subsection{T\lowercase{e}V $\gamma$-rays}

Since isotropy is implied in Eq.~(\ref{Qpi}), the $\gamma$-ray
emissivity is obtained from the $\pi^0$ emissivity as
\begin{equation}
Q_{\gamma}^{Ap} (E_\gamma) = 2 \, \int_{E_{\pi^0}^{\rm min}
  (E_\gamma)}^{E_{\pi^0}^{\rm max} (E_{n}^{\rm max})} \, \frac{Q_{\pi^0}^{Ap}
  (E_{\pi^0})}{\left(E_{\pi^0}^2 - m_{\pi^0}^2 \right)^{1/2}} \,
  dE_{\pi^0}
\end{equation}
where $E_{\pi^0}^{\rm min} (E_\gamma) = E_\gamma +
m_{\pi^0}^2/(4E_\gamma)$. Hence, the $\gamma$-ray emissivity is given
by
\begin{equation}
Q_{\gamma}^{Ap} (E_\gamma) \simeq Z_{\pi^0 \gamma} (\alpha) \,
Q_{\pi^0}^{Ap} (E_\gamma)
\end{equation}
with $Z_{\pi^0 \gamma} (\alpha) = 2 / \alpha$.

On the other hand, the $\gamma$-ray emissivity due to nuclei
photo-disintegration, Eq.~(\ref{qgdis}), in the range of energies where
$R_A$ is approximately constant, can be written as
\begin{equation}
Q_\gamma^{\rm dis} (E_\gamma) \simeq Z_{A \gamma} (\alpha) \, Q_A^{\rm dis}
(E_\gamma)
\end{equation}
where, assuming dominance of a single component (silicon),
\begin{eqnarray}
Z_{A \gamma} (\alpha) & = &
\left(\frac{\overline{n_{28}}}{\alpha}\right) \,
\left(\frac{2\overline{E'_{\gamma 28}}}{m_N}\right)^{\alpha - 1}
\\[2ex]
Q_A^{\rm dis} (E_N) & = & R_A \, \frac{dn}{dE_N} (E_N)
\end{eqnarray}

Hence, the ratio of the $\gamma$-ray emissivity due to these two
mechanisms is given by,
\begin{widetext}
\begin{equation}
R_{Ap/{\rm dis}}^{\gamma} (\alpha) \equiv \frac{Q_{\gamma}^{Ap}
  (E_\gamma)}{Q_\gamma^{dis} (E_\gamma)}  \simeq  0.1 \,
  \left(\frac{n_{\rm H}}{0.1 \, {\rm cm}^{-3}}\right) \,
\left(\frac{A}{28}\right)^{3/4} \,
\left(\frac{10^{-5} \, {\rm yrs}^{-1}}{R_A}\right) \,
\left(\frac{0.005}{Z_{A \gamma} (\alpha)}\right) \,
\left(\frac{Z_{A \pi^0} (\alpha) \, Z_{\pi^0 \gamma} (\alpha)}{Z_{A
    \pi^0} (2) \, Z_{\pi^0 \gamma} (2)}\right)
\label{idontknow}
\end{equation}
\end{widetext}
Thus, for $n_{\rm H} = 0.1 \, {\rm cm}^{-3}$ as quoted in the model of
Ref.~\cite{TDR04}, $\gamma$-rays from nuclei dissociation dominate
those from pion decays produced in hadronic interactions by as much as an
order of magnitude. Note however that in the model of
Ref.~\cite{TDR04} the volume-integrated luminosity $\int
Q_{\gamma}^{Ap} (E_\gamma) dV \propto n_{\rm H}^{-1/2}$. Thus, within this
model, a higher density would {\it decrease} the hadronic $\gamma$-ray
flux coming from this region. As suggested in Ref.~\cite{TDR04}, we
have taken $n_{\rm H} = 0.1 \, {\rm cm}^{-3}$ which, for the concentration
of stars in the Cygnus OB2 region, corresponds to a volume of
interaction equal to that of the region of the HEGRA signal.

Hence, $\gamma$-rays produced by the de-excitation of nuclei after
photo-dissociation in the photon background might well be the primary
contribution to the TeV signal from the Cygnus OB2 region.

\subsection{T\lowercase{e}V neutrinos}

The interaction of high energy nuclei with protons of the ISM gives
rise to a neutrino (and antineutrino) emissivity at production which,
in an analogous manner as in the previous case, can be expressed as
\begin{eqnarray}
\label{qnuAp}
Q_{\nu_\mu}^{Ap} (E_{\nu_\mu}) & = & 2 \, [Z_{\pi^0 \nu_\mu} (\alpha) +
  Z_{\mu \nu_\mu} (\alpha)] \, Q_{\pi^0}^{Ap} (E_{\nu_\mu}) \\[2ex]
Q_{\nu_e}^{Ap} (E_{\nu_e}) & = & 2 \, Z_{\mu \nu_e} (\alpha) \,
  Q_{\pi^0}^{Ap} (E_{\nu_e})
\end{eqnarray}
where the factor 2 takes into account the sum of both neutrino and
antineutrino emissivities, and $Z_{\pi^0 \nu_\mu} (\alpha)$, $Z_{\mu
  \nu_\mu} (\alpha)$ and $Z_{\mu \nu_e} (\alpha)$ are the $Z$-factors
corresponding to $\nu_\mu$ from pion and muon decay and $\nu_e$ from
muon decay, respectively. They are given by~\cite{KAB06}
\begin{eqnarray}
Z_{\pi^0 \nu_\mu} (\alpha) & =  & \frac{(1 - r)^{\alpha-1}}{\alpha}
\\[2ex]
Z_{\mu \nu_\mu} (\alpha) & =  & \frac{4 \, \left[3 - 2 \, r - r^\alpha
  \, (3 - 2 \, r + \alpha - \alpha \, r)\right]}{\alpha^2 \, (1 - r)^2
  \, (\alpha + 2) \, (\alpha + 3)} \\[2ex]
Z_{\mu \nu_e} (\alpha) & = & \frac{24 \, \left[\alpha \, (1 - r) - r
    \, (1 - r^{\alpha})\right]}{\alpha^2 \, (1 - r)^2 \, (\alpha + 1)
  \, (\alpha + 2) \, (\alpha + 3)}
\end{eqnarray}
where $r = (m_\mu/m_\pi)^2 = 0.467$. Hence, after taking into account
oscillations, the ratio of muon and electron neutrinos plus
antineutrinos to $\gamma$-rays of the same energy for $\alpha \in
(1.5,3)$ is approximately given by (and in good agreement with
Refs.~\cite{KHSA06,Gaisser,CV05,Anchordoqui:2004eu,Kistler:2006hp})
%
\begin{eqnarray}
R_{\nu_\mu/\gamma}^{A p} (\alpha) &
\sim & 1.05 - 0.28 \,\alpha \\[2ex]
R_{\nu_e/\gamma}^{A p} (\alpha) &
\sim & 1.08 -0.28 \,\alpha
\end{eqnarray}
%
Thus, the predicted ratios for $\nue$:$\numu$:$\nutau$:$\gamma$ at a common energy
are $\sim$1:1:1:1; here we have included $\numu$-$\nutau$ equilibration which gives
the added information that $\nutau$:$\numu$=1:1.

Note that at the high energies we are interested in, the ratio of the
number of neutrinos to that of $\gamma$-rays at the same energy, is
smaller than the expected 3:1. The reason being that the average
neutrino energy is about a quarter of that of the parent pion, while
in the case of the photon it is half. Hence there are more very low
energy neutrinos than very low energy photons, which compensates what
happens at higher energies, so that the ratio of the {\it total}
number of neutrinos to that of photons is indeed 3:1.

On the other hand, for the nuclei disintegration case, from
Eq.~(\ref{qnudis}), the neutrino emissivity at production can be
written as
\begin{equation}
Q_{\overline{\nu}_e}^{\rm dis} (E_{\overline{\nu_e}}) =
R_{n/\gamma}^{dis} (\alpha) \, Z_{n \nu} (\alpha) \,
Q_\gamma^{\rm dis} (E_{\overline{\nu}_e})
\end{equation}
where the $Z$-factor associated to the antineutrino emission from
neutron decay and the ratio of neutrons to $\gamma$-rays of the same
energy, $R_{n/\gamma}^{\rm dis}$, are given by
\begin{eqnarray}
Z_{n \nu} (\alpha) & = & \left(\frac{1}{\alpha}\right) \,
  \left(\frac{2 \, \epsilon_0}{m_N}\right)^{\alpha - 1} \\[2ex]
R_{n/\gamma}^{\rm dis} (\alpha) & = & \frac{1}{2 \, \overline{n_{28}}} \,
  \left(\frac{m_N}{\overline{E'_{\gamma 28}}}\right)^{\alpha-1}
\end{eqnarray}

Hence, the ratios of neutrinos (after taken into account oscillations)
due to the $Ap$ and photo-dissociation mechanisms are given by
\begin{widetext}
\begin{eqnarray}
R_{A p/{\rm dis}}^{\nu_\mu} (\alpha) & \simeq & 4.2 \, \left(\frac{n_{\rm H}}{0.1
  \, {\rm cm}^{-3}}\right) \, \left(\frac{A}{28}\right)^{3/4} \,
  \left(\frac{10^{-5} \, {\rm yrs}^{-1}}{R_A}\right) \,
  \nonumber \\
 & \times & \left(\frac{R_{\nu_\mu/\gamma}^{A p}
  (\alpha)}{R_{\nu_\mu/\gamma}^{A p} (2)}\right) \,
  \left(\frac{R_{n/\gamma}^{dis} (2) \, Z_{n \nu} (2) \, Z_{A \gamma}
  (2)}{R_{n/\gamma}^{dis} (\alpha) \, Z_{n \nu} (\alpha) \, Z_{A
  \gamma} (\alpha)}\right) \, \left(\frac{Z_{A \pi^0} (\alpha) \,
  Z_{\pi^0 \gamma} (\alpha)}{Z_{A \pi^0} (2) \, Z_{\pi^0 \gamma}
  (2)}\right) \\[2ex]
R_{A p/{\rm dis}}^{\nu_e} (\alpha) & \simeq & 1.8 \, \left(\frac{n_{\rm H}}{0.1 \,
  {\rm cm}^{-3}}\right) \, \left(\frac{A}{28}\right)^{3/4} \,
  \left(\frac{10^{-5} \, {\rm yrs}^{-1}}{R_A}\right) \,
  \nonumber \\
& \times & \left(\frac{R_{\nu_e/\gamma}^{A p} (\alpha)}{R_{\nu_e/\gamma}^{A p}
  (2)}\right) \, \left(\frac{R_{n/\gamma}^{dis} (2) \, Z_{n \nu} (2)
  \, Z_{A \gamma} (2)}{R_{n/\gamma}^{dis} (\alpha) \, Z_{n \nu}
  (\alpha) \, Z_{A \gamma} (\alpha)}\right) \, \left(\frac{Z_{A \pi^0}
  (\alpha) \, Z_{\pi^0 \gamma} (\alpha)}{Z_{A \pi^0} (2) \, Z_{\pi^0
  \gamma} (2)}\right)
\end{eqnarray}
\end{widetext}
i.e., $\numu$ (and $\nutau$) is suppressed by the factor 4.2 in
photo-dissociation relative to $Ap$, and $\nue$ is suppressed by 1.8.
We note that the neutrino flavor results here agree with the
well-known neutrino flavor ratios $\sim$ 5:2:2 and $\sim$ 1:1:1 from
astrophysical neutron and pion decay.

Therefore, if the TeV $\gamma$-ray signal from the Cygnus OB2
region can be explained by de-excitation of daughter nuclei after
photo-dissociation, the neutrino population will be, in general,
dominated by pion decay.  However, with a significant
population of $^4$He, this may not be the case: through
photo-dissociation $^4$He only contributes to the
$\overline{\nu_e}$-flux without affecting the photon signal.
Photon production from $^4$He $p$ interactions are negligible, see
Eq.~(\ref{idontknow}).

\section{E\lowercase{e}V cosmic ray excess?}
\label{VI}

A seemingly different, but in fact closely related subject we will
discuss in this section is the intriguing anisotropy that has been
observed by several experiments in the energy range near $10^9~{\rm
  GeV} \equiv {\rm EeV}$. Analyses of the data collected by the Akeno
Giant Air Shower Array (AGASA) seem to indicate a $4\sigma$ excess of
events from the direction of the Galactic plane in the narrow energy
window $10^{8.9}$~GeV to $10^{9.5}$~GeV~\cite{AGASA}. A Galactic plane
enhancement at the $3.2\sigma$ level between $10^{8.3}$~GeV and
$10^{9.5}$~GeV has also been reported by the Fly's Eye
Collaboration~\cite{FE}. The anisotropy in the AGASA data sample is
dominated by hot spots near the Galactic center and the Cygnus region.
However, based on a data sample larger than any previous experiment,
the Pierre Auger Collaboration has reported no evidence for such an
excess from the direction of the Galactic center~\cite{Auger}.

The complete isotropy below about $10^{7.7}$~GeV revealed by KASCADE
data~\cite{Antoni:2003jm} vitiate direction-preserving photons as
primaries.  Therefore, the excess from the Galactic plane is very
suggestive of neutrons as candidate primaries, because the directional
signal requires relatively-stable neutral primaries, and time-dilated
neutrons can reach the Earth from typical Galactic distances when the
neutron energy exceeds $10^{9}$~GeV. The analysis
of Fly's Eye data~\cite{FE} implies that if neutrons are the carriers
of the anisotropy, there {\em needs to be} some contribution from at
least one source closer than $\sim 2$ kpc.  As we mentioned in the
Introduction, the full Fly's Eye data include a directional signal
from the Cygnus region which was somewhat lost in an unsuccessful
attempt~\cite{Cassiday:kw} to relate it to $\gamma$-ray emission from
Cygnus X-3. This signal was also observed with the Akeno air shower
array (see Fig.~\ref{cygnusx3})~\cite{Teshima:1989fc}. As shown in
Fig.~\ref{cygOB2}, Cygnus OB2 overlaps along our line of sight with
Cygnus X-3, which lies about 8 kpc farther away than the stellar
association.

In the context of Cygnus X-3, we note that the 1989
analysis~\cite{Lawrence:1989xv} of data collected in the Haverah Park
experiment was consistent with the absence of anisotropy in the
direction of that source, in the energy bin of interest.  There was
indeed an excess, with a Poissonian probability of 0.013, but at
energies above $10^{9.6}$ GeV. However, a more recent analysis by the
group at Leeds~\cite{Ave:2001hq}, based on improved shower
simulations, more selective trigger criteria, and a reduction of the
zenith acceptance from 60$^\circ$ to $45^\circ$, reveals that
previous~\cite{Lawrence:1991cc} energy estimates need to be reduced by
about 30\%. It is then suggestive that the excess may overlap the
region of interest. One can encourage an event-by-event analysis to
see whether this is indeed the case; in the meantime one may reserve
judgment with respect to presence or absence of anisotropy in the
Haverah Park data.

\begin{figure}
\begin{center}
\postscript{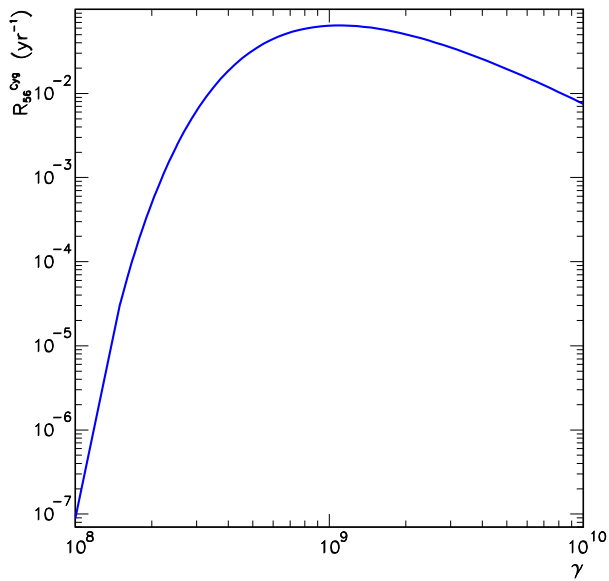}{0.98}
\caption{Photo-disintegration rate of $^{56}$Fe on the $T = 50~{\rm K}$
Cygnus OB2 blackbody radiation as given by Eq.~(\ref{RBE}). }
\label{Rcyg}
\end{center}
\end{figure}

More recently, the HiRes Collaboration has carried out a systematic
search for point sources in the Northern sky. In an initial series of
papers, they report no significant excess for energies greater than
$10^{9.5}$ GeV~\cite{Abbasi:2005mk}, $10^{10}$
GeV~\cite{Westerhoff:2005jh}, and $10^{10.6}$
GeV~\cite{Abbasi:2004vu}. All these energy thresholds exceed the range
above $10^{8.7}$~GeV where the Akeno and Fly's Eye anisotropies were
reported. Motivated by the present study, we have asked the HiRes
Collaboration to extend their analysis down to the energy threshold of
interest, in the direction of Cygnus. The result is that there is no
significant excess above background for the same angular and energy
cuts used by Akeno and Fly's Eye, this with a larger data sample than
available to either of these previous
studies~\cite{Stefan}. Additional scans were performed for nearby
energy and angular cuts, with the same result of no significant
excess. If the previously-reported anisotropies were not statistical
fluctuations, then they can be associated with bursts over periods
of less than 15 years. This possibility can be monitored with data to
be collected at the Northern site of the Pierre Auger
Observatory~\cite{Abraham:2004dt}.

In what follows, we will proceed on the assumption that the anisotropy
reported by the Fly's Eye~\cite{Cassiday:kw} and
Akeno~\cite{Teshima:1989fc} collaborations originated in a burst
taking place in Cygnus OB2 during a 10-yr period, and assess the
implications for source energetics.

For nuclei with energies EeV/nucleon, the dissociation proceeds via
collisions with far infrared thermal photons populating molecular
clouds with temperature $\sim 50~{\rm K}$~\cite{Wilson:1997ud}.  The
photo-disintegration rate is shown in Fig.~\ref{Rcyg}, giving an
interaction time of about 100~yr, which allows a perturbative
treatment during a 10 yr burst. We can then use
Eq.~(\ref{diegool})with the left hand side set equal to the neutron
luminosity for energies above $E_1 = 0.5~{\rm EeV}$. Specifically,
\begin{eqnarray}
\frac{dN}{dt} & = & 4 \pi d^2\,\,e^{d/4.6~{\rm kpc}}\,
\int_{E_1}^\infty  \left.
\frac{dF_n}{dE_n}\right|_\oplus \, dE_n \nonumber \\
 & \approx & 7 \times 10^{27} \,\,\, {\rm neutrons} \, {\rm s}^{-1}
\,\,,
\end{eqnarray}
where
\begin{equation}
\int_{E_1}^\infty \left. \frac{dF_n}{dE_n}\right|_\oplus \,
dE_n \simeq 2 \times 10^{-17}~{\rm cm}^{-2}\,\,{\rm s}^{-1}
\end{equation}
is the flux observed at Earth~\cite{Cassiday:kw,Teshima:1989fc}, and
the factor $e^{d/4.6~{\rm kpc}}$ corrects for neutron
$\beta$-decay. Now, taking $R^{\rm Cyg}_{56} \approx 10^{-2}~{\rm
  yr}^{-1}$ we obtain
\begin{equation}
\int_{E_1} \frac{dn_{\rm Fe}}{dE_N}\, dE_N = 5.6
\times 10^{-30}~{\rm cm}^{-3}\, .
\label{J}
\end{equation}
Assuming a nucleus power law spectrum $dF_{\rm Fe}/dE = N_{56}\,
E^{-2}$, Eq.~(\ref{J}) will determine the normalization
constant. Then, the energy density can be obtained via
\begin{equation}
{\cal E} = \int_{E_1} E\,\, \frac{dn_{\rm Fe}}{dE}\, dE \,,
\end{equation}
and so straightforward calculation shows that the total energy
$4\pi R^3 {\cal E}/3 = 5 \times 10^{35}~{\rm erg}$ is well within the
kinetic energy budget of a 10~yr burst.

\section{Conclusions}
\label{VII}

Over the last few decades, the Cygnus region has been the focus of
extensive study with a view to understanding astrophysical phenomena
in very energetic environments. Interesting observations have been
made, but have not held up at statistically significant levels. More
recently, improved efficiency and higher resolution have provided
signals at discovery level. Perhaps, the most precise of such signals
is the (1999-2002) HEGRA detection of the unidentified TeV J2032+4130
source, with an average flux $\sim 3\%$ of the Crab
Nebula~\cite{HEGRA05}. This bright ``hot spot'', at the edge of the
very active Cygnus OB2 star association, has been recently confirmed
through a re-analysis of ``old'' data collected with the Whipple
Observatory~\cite{Lang:2004bk}.  However, the average flux ($\sim 12\%$ of
the Crab) detected during 1989-1990 is well above that reported by the
HEGRA Collaboration.  Neither the Whipple nor the HEGRA experiments
see any evidence for variability within their individual databases.
The large differences between the flux levels cannot be explained as
errors in estimation of the sensitivity of the experiments because
they have been calibrated by the simultaneous observations of other
TeV sources.  Certainly more data are needed to resolve this issue.
Happily, data are currently flowing from the Milagro
telescope~\cite{Goodman} and the Tibet Air Shower
Array~\cite{Amenomori:2006bx} experiments at a significant rapid rate.

In this paper, we have focussed on the unidentified TeV $\gamma$-ray
source observed by the HEGRA Collaboration. Because of the absence of
X-ray or radio counterparts, it has been difficult to provide a
compelling explanation for the origin of the $\gamma$-rays from the
HEGRA source. This has motivated us to present an alternative model,
which has some well-defined predictions for future observations, both
in $\gamma$-ray facilities and at neutrino telescopes such as IceCube.
The model proposes that the observed TeV $\gamma$-rays are the result
of Lorentz-boosted MeV $\gamma$-rays emitted on the de-excitation of
daughter nuclei following collisions of PeV nuclei with a hot
ultraviolet photon background. Our results include the following:

(1) There is a specific prediction of a suppression on the
$\gamma$-ray spectrum in the region $\alt 1~{\rm TeV}$ -- this because
of the need to achieve the Giant Dipole Resonance for disintegration
through collision with $\sim$ few eV photons.

(2) The flux of $\gamma$-rays produced through nuclei de-excitation
process  dominates by an order of magnitude that resulting from
$\pi^0$ production and decay in collisions of nuclei with the
ambient gas background.

(3) The presence of a significant neutrino signal from decay of
neutrons produced in the course of the photo-disintegration is
strongly contingent on the abundance of $^4$He in the source,
mirroring its observed prevalence in cosmic rays. This compensates
by an order of magnitude over its reduced photo-disintegration
rate, and results in a $\overline \nu_\mu$ signal at IceCube
(after oscillation from $\overline \nu_e$) comparable to the
$\nu_\mu$ signal from hadronic interactions (as noted previously,
this is sufficient to establish a 15$\sigma$ discovery in 15 years
of observation.)The ability of IceCube to measure flavor ratios
can distinguish between these possibilities~\cite{Beacom:2002vi}.

(4) For this model to succeed, a high efficiency is required for
accelerating a low energy population of nuclei at the source to PeV
energies. Thus, evidence for the model (such as a substantial
suppression in the $\gamma$-ray spectrum below $\sim 1~{\rm TeV}$)
could provide some insight into source dynamics.

(5) It is noteworthy that the contribution to the diffuse $\gamma$-ray
flux component resulting from the interaction of extreme-energy cosmic
ray nuclei with the background radiation fields permeating the
universe~\cite{CIRB} is well below the observed EGRET
data~\cite{Sreekumar:1997un} (for details see Ref.~\cite{shortP}).

In summary, the availability of new data in coming years in
$\gamma$-ray, X-ray, neutrino and cosmic-ray sector experiments affords an
outstanding opportunity for study of high energy processes in
extreme environments. Careful exploration of model implications
will provide  an important complementary role to these
observations.

\acknowledgments{We would like to thank Felix Aharonian, Gavin Rowell,
  and the HEGRA Collaboration for permission to reproduce
  Fig.~\ref{cygOB2}. We would also like to thank Tom Paul, Subir
  Sarkar and Stefan Westerhoff for valuable discussions on the EeV
  (an)isotropy. We are indebted to the HiRes Collaboration for
  permission to make public their results on the anisotropy search
  around the Cygnus region.  We are grateful to Rene Ong for comments
  and a careful reading of the manuscript. The research of LAA is
  supported by the University of Wisconsin-Milwaukee. JFB is supported
  by The Ohio State University and the NSF CAREER Grant No.
  PHY-0547102. HG is supported by the U.S. National Science Foundation
  (NSF) Grant No PHY-0244507. SPR is supported by the NASA Grant
  ATP02-000-0151 and by the Spanish Grant FPA2005-01678 of the MCT.
  TJW is supported by the U.S. Department of Energy Grant No.
  DE-FG05-85ER40226, by the NASA Grant ATP02-000-0151, and by a
  Vanderbilt University Discovery Award.}

\section*{Note added}
\noindent In the accompanying {\it Letter}~\cite{shortP} we have
adopted a more descriptive notation for some variables. Throughout
this paper we use the traditional notation, which shortens the text
length of the equations.  The equivalences are: $\gamma = \Gamma_A, \
\ \epsilon' = \Gamma_A \epsilon, \ \ \ \ \epsilon_0' =
\epsilon_\gamma^{\rm GDR}, \ \ \overline{E'_{\gamma A}} =
\epsilon_\gamma^{\rm dxn}, \ E_\gamma = \epsilon_\gamma^{\rm LAB}, \ E
= E_{\rm A}^{\rm LAB}, \ \Gamma = \Gamma^{\rm GDR},$ and $\sigma_0 =
\sigma^{\rm GDR}.$


\begin{thebibliography}{99}
\bibitem{note94}
  The factor $9/(4 R^2)$ is the result of calculating $\int d^3\vec r_1\ \
  \int d^3\vec r_2\ \ |\vec r_1 - \vec r_2|^{-2}\ (4\pi R^3/3)^{-2},$
  where $\vec r_1$ is the position of a star and $\vec r_2$ is the
  position of an observer (the position of the reaction) in a region
  $R$ with uniformly distributed sources.
\bibitem{note94b} The $47/(4 R_{\rm out}^2)$ factor is the
  result of calculating $\ \int d^3\vec r_1\ \ \int d^3\vec
  r_2\ \rho(r_1) \ |\vec r_1 - \vec r_2|^{-2} \ (N_O + N_B)^{-1}
  (4\pi r^3/3)^{-1},$ where $\rho(r_1)$ is the density of stars in a
  region of radius $R_{\rm out}$ and the reaction occurs in a region of radius
  $r$.
\end{thebibliography}
\end{document}